\documentclass[fleqn,usenatbib]{mnras}

\usepackage{newtxtext,newtxmath}
\usepackage[dvipsnames]{xcolor}
\usepackage[pagewise]{lineno}
\usepackage[T1]{fontenc}
\usepackage{natbib}
\usepackage{orcidlink}
\DeclareRobustCommand{\VAN}[3]{#2}
\let\VANthebibliography\thebibliography
\def\thebibliography{\DeclareRobustCommand{\VAN}[3]{##3}\VANthebibliography}

\usepackage{graphicx}	% Including figure files
\usepackage{amsmath}	% Advanced maths commands

\usepackage{booktabs} % For professional-looking tables
\usepackage{caption}  % For table captions
\usepackage{array}    % For better column formatting

\newcommand{\red}{\textcolor{black}}
\title[Semi-stochastic corrections for galaxy models]{Optimised neural network predictions of galaxy formation histories using semi-stochastic corrections}

\author[Behera, Tojeiro \& Chittenden 2025]{
Jayashree Behera$^{\orcidlink{0009-0002-2434-5903} \ 1}$,
Rita Tojeiro$^{\orcidlink{0000-0001-5191-2286} \ 2}$\thanks{E-mail: rmftr@st-andrews.ac.uk}
and Harry George Chittenden$^{\orcidlink{0000-0001-5856-8713} \ 2,3}$
\\
$^{1}$Department of Physics, Kansas State University, Cardwell Hall, 116, 1228 N M.L.K. Jr. Dr, Manhattan, KS 66506, United States\\
$^{2}$School of Physics \& Astronomy, University of St Andrews, North Haugh, St Andrews KY16 9SS, Scotland, United Kingdom\\
$^{3}$Centre for Astrophysics and Supercomputing [CAS], Swinburne University of Technology, P.O. Box 218, Hawthorn VIC 3122, Melbourne, Australia
}

\date{Accepted XXX. Received YYY; in original form ZZZ}

\pubyear{2025}

\begin{document}
\label{firstpage}
\pagerange{\pageref{firstpage}--\pageref{lastpage}}
\maketitle

\begin{abstract}
We present a novel methodology to improve predictions of galaxy formation histories by incorporating semi-stochastic corrections to account for short-timescale variability. Our paper addresses limitations in existing models that capture broad trends in galaxy evolution, but fail to reproduce the bursty nature of star formation and chemical enrichment, resulting in inaccurate predictions of key observables such as stellar masses, optical spectra, and colour distributions. We introduce a simple technique to add a stochastic components by utilizing the power spectra of galaxy formation histories. We justify our stochastic approach by studying the correlation between the phases of the halo mass assembly and star-formation histories in the IllustrisTNG simulation, and we find that they are correlated only on timescales longer than 6 Gyr, with a strong dependence on galaxy type. \red{We demonstrate our approach by applying our methodology to the predictions on a neural network trained on hydrodynamical simulations, which failed to recover the high-frequency components of star-formation and chemical enrichment histories. Our methodology }successfully recovers realistic variability in galaxy properties at short timescales. It significantly improves the accuracy of predicted stellar masses, metallicities, spectra, and colour distributions and provides a practical framework for generating large, realistic mock galaxy catalogs, while also enhancing our understanding of the complex interplay between galaxy evolution and dark matter halo assembly.

\end{abstract}

\begin{keywords}
galaxies: evolution, galaxies: formation, galaxies: haloes, galaxies: star formation
\end{keywords}

%%%%%%%%%%%%%%%%%%%%%%%%%%%%%%%%%%%%%%%%%%%%%%%%%%

%%%%%%%%%%%%%%%%% BODY OF PAPER %%%%%%%%%%%%%%%%%%

\section{Introduction}

According to the current model of galaxy evolution and structure formation, galaxies are formed by the contraction of baryonic gas that is gravitationally bound to a dark matter halo. The evolution of a galaxy is therefore expected to be at least partly determined by the properties and evolution of its halo and surrounding environment. Although this relationship has not yet been fully understood, studies in simulations and observations point to strong links between the two. For example, in cosmological hydrodynamical simulations, the halo formation history has been shown to impact on a galaxy's present-day stellar mass (e.g. \citealt{2023MNRAS.518..562A, 2021NatAs...5.1069C}), star-formation rate or colour (e.g. \citealt{2021MNRAS.508..940M}), circumgalactic medium \cite{2021MNRAS.501..236D}, and morphology (e.g. \citealt{2021MNRAS.501..236D}), at fixed stellar or halo mass or using controlled experiments. Although there is broad agreement that halo growth and galaxy growth are linked, we still lack clarity in the details with different simulations and observations arguing for different - sometimes directly opposing - effects of how halo assembly histories impact galaxy properties. The difficulty comes from the complexity of this galaxy-halo connection and its dependence on physical processes that span decades in physical and time scales, as well as the interaction between baryons and dark matter. In observations, we remain limited by the fact that halo properties - and halo assembly in particular - remain notoriously hard to measure.

Recent machine learning techniques have shown promise in emulating complex astrophysical processes that connect the properties of dark matter halos to those of the galaxies they host. These efforts have two broad goals: one is to enable the creation of much larger mocks with realistic galaxy populations than what is possible with hydrodynamical simulations, in a fast and efficient way; and another is to study the impact of different halo properties on the build-up of the galaxy-halo connection. And, indeed, such efforts have been shown to offer a way to efficiently populate large N-body simulations with mock galaxy catalogs while encapsulating key relationships between halos and galaxies \citep{2018MNRAS.478.3410A,2019MNRAS.489.3565J,2019arXiv191007813Y, 2020arXiv201200111W}. 

In a recent study, \cite{2023MNRAS.518.5670C} (CT23 henceforth) developed neural networks to predict the star formation and chemical enrichment histories of central and satellite galaxies based solely on the properties of their dark matter halos and environments. The networks were trained on hydrodynamical simulations from the IllustrisTNG project \citep{2018MNRAS.480.5113M, 2018MNRAS.475..676S, 2018MNRAS.475..624N, 2019ComAC...6....2N, 2018MNRAS.473.4077P} and successfully recovered key observational benchmarks like the stellar mass-halo mass relation, mass-metallicity relation, and colour bimodality. The model utilizes a semi-recurrent neural network (rNN) algorithm to predict the time-resolved star formation history (SFH) and metallicity (or chemical enrichmeny) history (ZH) of central and satellite galaxies from the historical evolution of their dark matter halos and local dark matter environment. From these properties, one can then self-consistently predict observables such as optical spectra and broadband colours. The model is fairly practical and can be used in high fidelity N-body simulations, statistically reproducing key relationships of the galaxy-halo connection, and producing realistic mock surveys of unprecedented size and complexity \citep{2024arXiv240916079C}.
While the CT23 networks broadly matched time-averaged trends, the predicted histories lacked variability on short timescales compared to the original hydrodynamic histories. \red{By correlating the lack of power on short timescales of the formation histories with residuals in predicted vs true galaxy properties, CT23 concluded that the lack of variablity on short timescales} led to a systematic underestimation of stellar masses, luminosities, and emission line strengths for some galaxies. \red{They also noted a systematic underestimation of the scatter in fundamental relations such as the mass-metallicity relation.} The discrepancy was more significant for star-forming central galaxies than quenched satellites.

Modelling and measuring short-timescale variability in star-formation histories and its connection to fundamental relations is an active area of research. Studies on cosmological hydrodynamical simulations have linked long timescales ($\gtrsim 1$ Gyr) with halo-related processes (e.g. accretion) and shorter timescales with variability associated with feedback processes (e.g. \citealt{2015MNRAS.447.3548S, 2020MNRAS.498..430I}). The lack of a direct physical link between halo and star-formation histories on short timescales then becomes a natural and tempting explanation for the shortcomings of the networks in CT23. Purely stochastic methods to model the star-formation histories of galaxies have been successful at describing the scatter of the star-forming main sequence \citep{2019MNRAS.487.3845C} and identifying different physical contributions to variability on different timescales by looking at different simulations of galaxy evolution (e.g. \citealt{2020MNRAS.498..430I}). Within these frameworks, for example, it is possible to explicitly model \red{ variability arising from different physical processes, such as gas inflow rates and the formation and destruction of giant molecular clouds,} as well as their impact on observables, laying the ground for observational constraints of these models (e.g. \citealt{2020MNRAS.497..698T, 2024ApJ...961...53I}).

In this follow-up paper to CT23, we implement a stochastic correction method to account for unmodelled short-timescale variability in the predicted star formation and metallicity histories. We have two main goals. One is to directly improve the predictions of the CT23 networks. Along with predicting the histories directly, we also train auxiliary neural networks to {\it directly} predict their power spectra as a function of halo properties. The power spectrum captures the degree of variability on different timescales, which can be attributed to physical drivers like stellar feedback, fluctuations in gas accretion, and mergers. This method utilizes randomized realizations of the histories by inverse Fourier transforming the predicted power spectra with quasi-random phases. This stochastic approach self-consistently introduces variability on short timescales while preserving the large-scale trends predicted by the original networks. In this paper we limit our corrections to timescales longer than 860 Myr, as that was the limiting frequency of the star-formation and chemical enrichment histories used by the neural-network. Future work will study the connection between halos and baryons using a larger range of frequencies. Our other goal is to understand the connection between halo assembly, star-formation and chemical enrichment histories by looking at the correlations in phases of these two components. Effectively, we ask the question: {\em on what timescales can the phases of star-formation and chemical enrichment histories be treated as purely stochastic, given the mass assembly history of a host halo?}

The added variability introduced by our quasi-stochastic corrections produces stellar masses, spectra and metallicities that show substantially better agreement with the reference hydrodynamic simulation, confirming that the missing short-timescale variability was a primary cause of the original discrepancies. 

Our paper is organised as follows. In Section~\ref{sec:data} we summarise the CT23 model and data, as well as the Fourier transforms data used here.  In Section~\ref{sec:corrections} we explore the phase correlations between star-formation, metallicity and halo mass assembly histories with the goal of informing the limits of our stochastic corrections, and we present our formalism and the details of their implementation. In Section~\ref{sec:results} we present our results and finally in Section~\ref{sec:conclusions} we discuss our results and conclude.

\section{Model and Data}\label{sec:data}

\begin{figure*}
    \includegraphics[width = \textwidth]{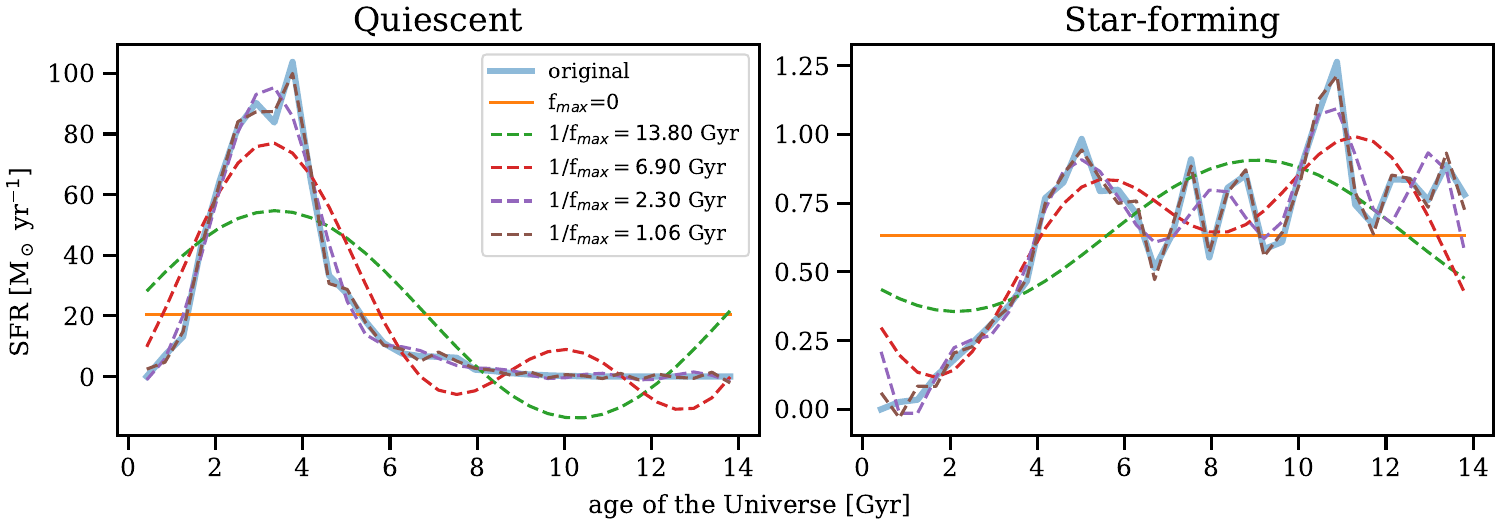}
    \caption{This figure demonstrates the decomposition of the star-formation history of two galaxies onto the frequency components of the DFT. In each panel, the thick blue line shows the original star formation history. The other lines show the reconstructed star-formation history using the frequencies up to (and including) the frequency listed on the label - we only use four frequencies here for clarity. The first component of the DFT is simply the mean the star-formation history over cosmic time, represented by the orange solid line. The left panel shows a typically quiescent high-mass galaxy ($M_* = 10^{11.2} M_\odot$) whereas the right panel shows a typically star-forming low-mass galaxy ($M_* = 10^{9.7} M_\odot$). }
    \label{fig:SFH_cen_rec}
\end{figure*}

\begin{figure*}
        \includegraphics[width = \textwidth]{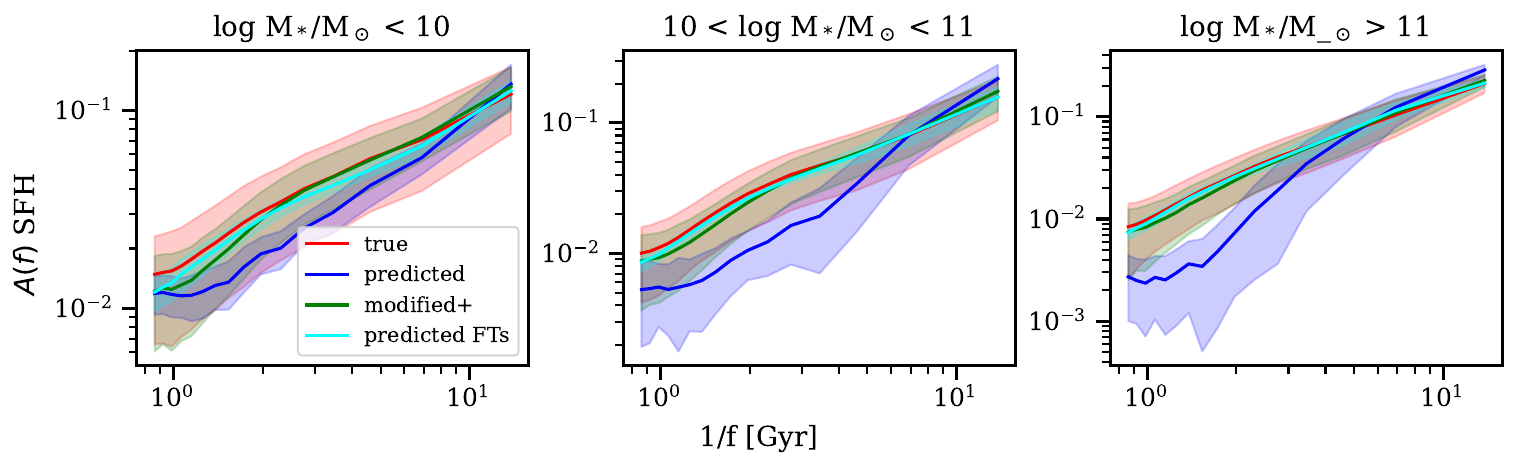}
        \includegraphics[width = \textwidth]{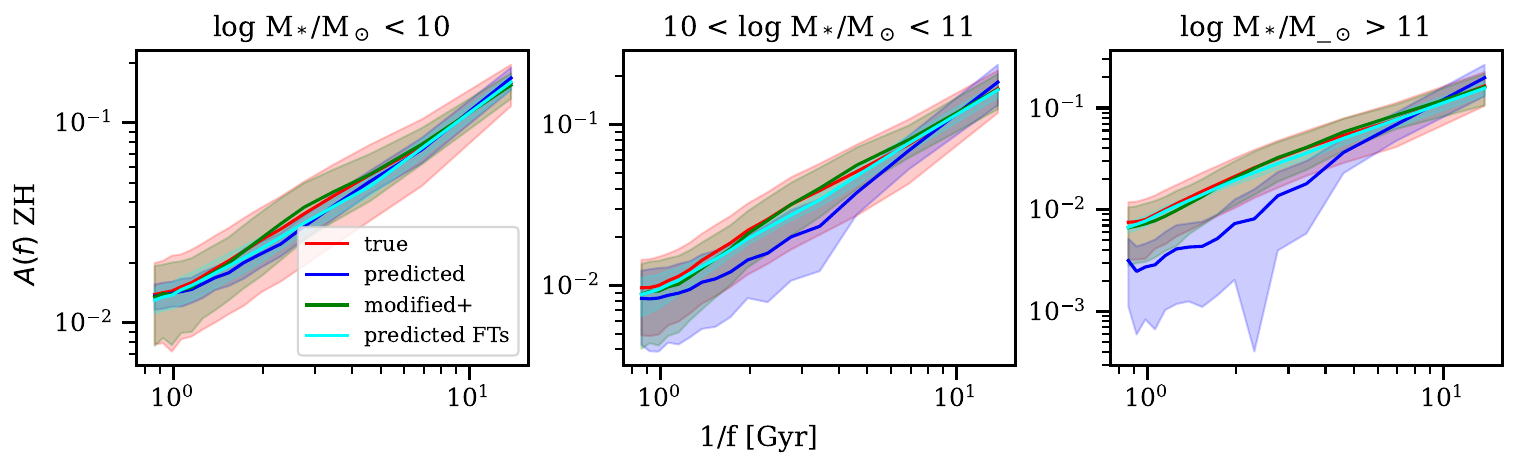}
    \caption{The mean amplitude of the DFT, $A(f)$, for the star-formation histories (top) and metallicity histories (bottom) of different samples of central galaxies. From left to right the different panels represent galaxies of low, intermediate, and high stellar mass, respectively. The different coloured lines show the mean of the true (red), predicted (blue), and modified+ (green) values. The cyan line shows the mean of the FFT amplitudes that were {\em directly} predicted by the rNN. The difference between the red and blue lines indicates the initial problem that this work aims to solve - the lack of power at small scales in the star-formation and metallicity histories predicted by the rNN (see Section~\ref{sec:data}). The similarity between the cyan and the red line motivates the approach proposed here - to use the predicted DFT amplitudes as an estimate of the true amplitudes (see Section~\ref{sec:corrections}). The difference between the green and red lines indicates that the methodology presented here is successful at recovering the power on all timescales (see Section~\ref{sec:results}). In all cases, the shaded areas show the standard deviation around the mean. }
    \label{fig:FTamp_cen}
\end{figure*}

We use the same SFHs, ZHs and \red{mass} assembly histories (MAHs) \red{of the dark matter halos} as in CT23, computed from the Illustris TNG simulation. SFHs and ZHs are defined from the mass-weighted age and mass-weighted metallicity distributions of star particles bound to each subhalo at z = 0, respectively. These properties are then binned on to a time grid that corresponds to every third snapshot of the TNG simulations, spanning a redshift range of 0 $\leq$ z $\leq$ 20. The final grid therefore has 33 bins, each corresponding to a lookback time interval of around 200 to 400 Myr. For full details on the neural network model, targets, features, and predictions, see CT23. Briefly, a combination of temporal features (such as the halo mass accretion rate history) and non-temporal features (such as the $z=0$ halo mass or infall velocity for satellites) are fed to a semi-recurrent neural network to predict time-dependent star-formation and chemical enrichment histories of galaxies in the same time grid. To increase the sampling of high-mass objects, CT23 combined data from TNG100 and TNG300 after correcting SFHs and ZHs for mass resolution effects.

For the neural-network modeling, the TNG data is split into training and test datasets. The training set consists of 75 percent of the combined TNG100 and resolution-corrected TNG300 data. Separate training sets exist for central and satellite galaxy neural networks. 20 percent of the training data is held out for validation during training iterations. The rest of the 25 percent the TNG galaxy data is reserved strictly for testing and is not used to update model parameters during training. Once trained, the neural networks are used to predict SFHs and ZHs from the halo properties of the test set. The predicted galaxy properties are compared with the true ones to evaluate to quantify model performance and check for overfitting. CT23 demonstrates that, on the whole, the recurrent neural network successfully recovers key properties of the galaxy population, namely the mean stellar to halo mass relation (SHMR, CT23-Fig 11) and the dependence of its scatter on halo assembly (CT23-Fig 15), stellar mass - magnitude relations (CT23-Fig 20), the relationship between galaxy SFH and halo mass (CT23-Fig 12) and its dependence of the stellar mass-weighted age with stellar mass (CT23-Fig 14). None withstanding these key successes, the neural network fails to recover the scatter in the mass-metallity relation (CT23-Fig 16), the scatter in galaxies' spectral energy distributions (SEDs, CT23-Fig17), or H$_\alpha$ luminosity. The network also clearly fails to predict SFH and ZH features on short time-scales (CT23-Fig13).

To analyse and mitigate this shortcoming of the rNN, we begin by comparing the amplitude of the Fourier transform (\textcolor{black}{FT}) of the true and predicted SF and ZHs. To do that, we compute the FT of each galaxy's SFH and ZH (and, later, MAH) using a NumPy\footnote{https://numpy.org} implementation of the discrete Fourier transform (DFT\footnote{https://numpy.org/doc/stable/reference/routines.fft.html}), allowing us to compare predicted and true histories as a function of time-scale. \textcolor{black}{We will represent the FTs of time-domain function \( h(t) \) as complex functions \( H(f) = A(f)e^{-i\theta(f)} \), where \( f \) denotes frequency, and \( A(f) \) and \( \theta(f) \) are the amplitude and phase as functions of frequency, respectively}. For each of our functions sampled in 33 bins in time (SFH, ZH, MAH), we obtain a DFT with 17 amplitudes and phases. The zeroth array element corresponds to the average function value, with phase $\theta=0$ by default. Each of the other 16 elements $i$ contains the DFT at a frequency that is $1/i^{th}$ of the period of the function, which is assumed to be age of the Universe at $z=0$ in the TNG cosmology. In other words, the frequency corresponding the $i=1$ frequency corresponds to a mode with a period $P=13.80$ Gyr, $i=2$ to a mode with a period of $P=6.90$ Gyr and so forth. The highest sampled frequency is $1.16$ Gyr$^{-1}$ equivalent to a timescale of $0.86$ Gyr.

For illustration purposes, we show in Fig~\ref{fig:SFH_cen_rec} the reconstruction of the \red{true} star-formation history of two galaxies \red{in our training sample} as a function of the maximum frequency (or, inversely, minimum timescale) used. Fig~\ref{fig:FTamp_cen} shows the average amplitude A$(f)$ of the SFHs and ZHs in three stellar mass bins that we will call "low mass" ($\log M_*/M_\odot < 10$), "intermediate mass" ($10 \leq \log M_*/M_\odot < 11$), and "high mass" ($\log M_*/M_\odot \ge 11$). The magnitude of $A(f)$ depends on the stellar mass of the galaxy, so amplitudes are normalised to unity for each galaxy prior to averaging, which allows us to compare the shape of A$(f)$ for the true (red) and predicted (blue) SFH and ZHs. In the case of the SFHs (top) we can see the clear deficit in power of the SFHs \red{predicted by the CT23 rNN (blue line)} on timescales shorter than approximately 5 Gyr, particularly at high stellar mass. In the case of the ZHs (bottom) this deficit is not seen in the lowest stellar mass bin. 

\begin{figure*}
    \includegraphics[width = \textwidth]{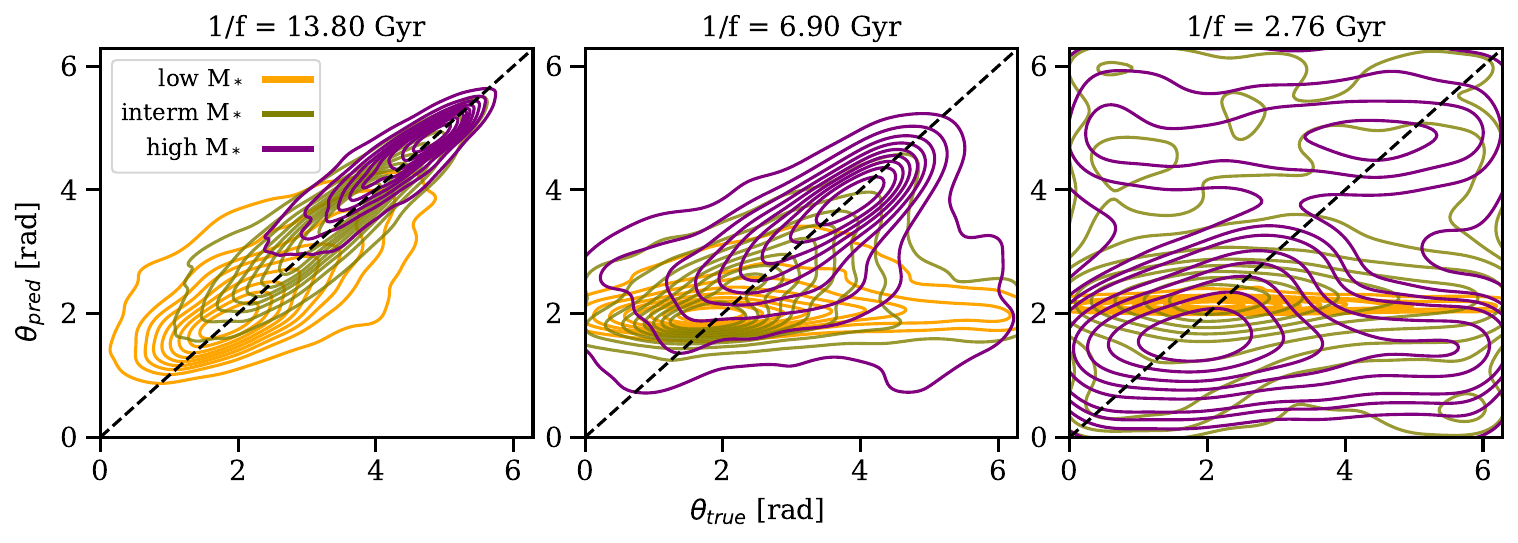}
    \caption{The phases of the predicted star-formation histories vs the phases of the true star-formation histories of central galaxies shown at three timescales: 13.8 Gyr (left), 6.90 Gyr (middle), and 2.76 Gyr (right). \red{Each panel shows a kernel-density estimation visualisation of the phases for galaxies in the three mass bins, with contours representing levels of equal probability density, in 10 equal bins}. The black dashed line is the one-to-one line. As we move to shorter timescales and to lower stellar masses, the correlations between true and predicted phases breaks down.}
    \label{fig:FTphases_cen_SFH}
\end{figure*}

We can also analyse how the phases of the SFH and ZHs \red{predicted by the rNN} compare to the true ones. Fig.~\ref{fig:FTphases_cen_SFH} shows the predicted vs true phases the SFHs of \red{central} galaxies in the test sample, at three different frequencies and in three mass bins \red{(the resulting figure for satellites is similar and not shown)}. Although the SFHs \red{predicted by the rNN} have phases that are well predicted on the lowest frequency, the correlation quickly deteriorates with increasing frequency, and phases for frequencies higher than $1/2.76$ Gyr$^{-1}$ are uncorrelated across true and predicted SFHs. We also note that the distribution of $\theta(f)$ is not uniform in $[0,2\pi]$, even for high frequencies - in other words, the range of allowed values of $\theta$ is determined by galaxy evolution processes, and carries physical meaning.
This can also be seen through the dependence on stellar mass, in that the phases of the more massive galaxies are both shifted with respect to those of lower mass galaxies (reflecting the earlier epoch of star formation) and are slightly better predicted by the rNN, particularly towards lower frequencies. We will investigate some of the reasons for this in Section~\ref{sec:phase correlations}. For now, we point out that the peak of the SFH or ZH of a galaxy is largely determined by the phase of the longest timescale mode (see e.g. Fig.~\ref{fig:SFH_cen_rec}) and that these are well recovered by the rNN at all stellar masses. However, the limitations towards higher frequencies are evident. In this paper we propose a solution to this problem by modifying the SFHs and ZHs \red{predicted by the rNN} such that they have the similar FTs to the true SFHs, \red{and we will refer to those as "modified/modified+", depending on the details of the modification.} \red{We will refer to quantities directly obtained from the TNG training samples as "true" and to the quantities predicted by the rNN as "predicted" (matching those in CT23).}

This line of investigation is possible because the rNN is able to predict {\em directly} the shape of A$(f)$ very well (i.e., A$(f)$ is explicitly made a target of the NN), although with a lower scatter than those corresponding to the true SFHs - see cyan line in Fig.~\ref{fig:FTamp_cen}. In the next section we introduce a formalism to correct the SFHs and ZHs in the case that the true amplitude of the FT is known (i.e., assumed to be correctly predicted by the rNN) and, in Section~\ref{sec:phase correlations}, we consider how to treat the phases in detail. 

\section{Semi-Stochastic corrections}\label{sec:corrections}

\subsection{Formalism}\label{sec:formalism}

We aim to modify the predicted star formation and \textcolor{black}{chemical enrichment (metallicity)} histories from our neural networks to better match the intrinsic variability of the original ones in the hydrodynamic simulations. We assume an additive relationship between the true and predicted histories and their respective FTs:
\begin{align}
    h_\text{true}(t) = h_\text{pred}(t) + h_\text{stoc}(t)
    \label{eq:sfh}
\end{align}
\begin{align}
    H_\text{true}(f) = H_\text{pred}(f) + H_\text{stoc}(f)
    \label{eq:fsfh}
\end{align}
where $h_\text{true}(t)$ and $h_\text{pred}(t)$ are the true and predicted histories in real time, and $h_\text{stoc}(t)$ is the correction term we seek to compute. \textcolor{black}{We denote the FT of the signal in the time domain $h_j(t)$ as $H_j(f) = \mathcal{F}[h_j(t)]$, where \( \mathcal{F} \) is the FT operator and \( \mathcal{F}^{-1} \) its inverse}. Each component in frequency space can be expressed in complex form as:
\begin{equation}
    H_j(f) = A_j e^{-i\theta_j} = A_j \cos{\theta_j} - i A_j \sin{\theta_j}, \quad j \in \{\text{true}, \text{pred}, \text{stoc}\}.
\end{equation}

Substituting into equation~\ref{eq:fsfh} and equating the real and imaginary parts, we obtain:
\begin{equation}
\begin{aligned}
    A_\text{true}\cos{\theta_\text{true}} &= A_\text{pred}\cos{\theta_\text{pred}} + A_\text{stoc}\cos{\theta_\text{stoc}}, \\
    A_\text{true}\sin{\theta_\text{true}} &= A_\text{pred}\sin{\theta_\text{pred}} + A_\text{stoc}\sin{\theta_\text{stoc}}.
\end{aligned}
\end{equation}

Solving these equations for $A_\text{stoc}$ and $\theta_\text{stoc}$ gives:
\begin{equation}
\begin{aligned}
    A_\text{stoc} &=  
    \left[ ( A_\text{true}\cos{\theta_\text{true}} - A_\text{pred}\cos{\theta_\text{pred}} )^2 \right. \\
    & \quad \left. + ( A_\text{true}\sin{\theta_\text{true}} - A_\text{pred}\sin{\theta_\text{pred}} )^2 \right] ^{1/2}, \\
    \theta_\text{stoc} &= \tan^{-1}\left( \frac{ A_\text{true}\sin{\theta_\text{true}} - A_\text{pred}\sin{\theta_\text{pred}} }{ A_\text{true}\cos{\theta_\text{true}} - A_\text{pred}\cos{\theta_\text{pred}} } \right).
\end{aligned}
\end{equation}

The amplitude, $A_\mathrm{stoc}$ represents the misfit between true and predicted transforms while the phase $\theta_\mathrm{stoc}$ is randomized. This defines the complete complex correction term $H_\mathrm{stoc}(f)$, that introduces missing stochasticity while preserving the predicted large-scale trends. We then obtain the correction in real space by taking the inverse FT:
\begin{equation}
    h_\text{stoc}(t) = \mathcal{F}^{-1}\big[H_\text{stoc}(f)\big].
\end{equation}

Finally, the corrected star formation or enrichment history is computed by adding this term to the predicted one, as in equation~\ref{eq:sfh}.

To determine the true amplitudes $A_\mathrm{true}$ and true phases  $\theta_\mathrm{true}$ we make the key assumption that the neural network accurately predicts the FT amplitudes (see cyan and red lines in Fig.~\ref{fig:FTamp_cen}), which govern the distribution of variability timescales. Under this assumption, we substitute the FT amplitudes directly predicted by the neural network as the true amplitudes. Consequently, the only remaining parameter to be determined is  $\theta_\mathrm{true}$.

\subsection{Phase correlations and stochasticity}\label{sec:phase correlations}

\begin{figure*}
    \includegraphics[width = \textwidth]{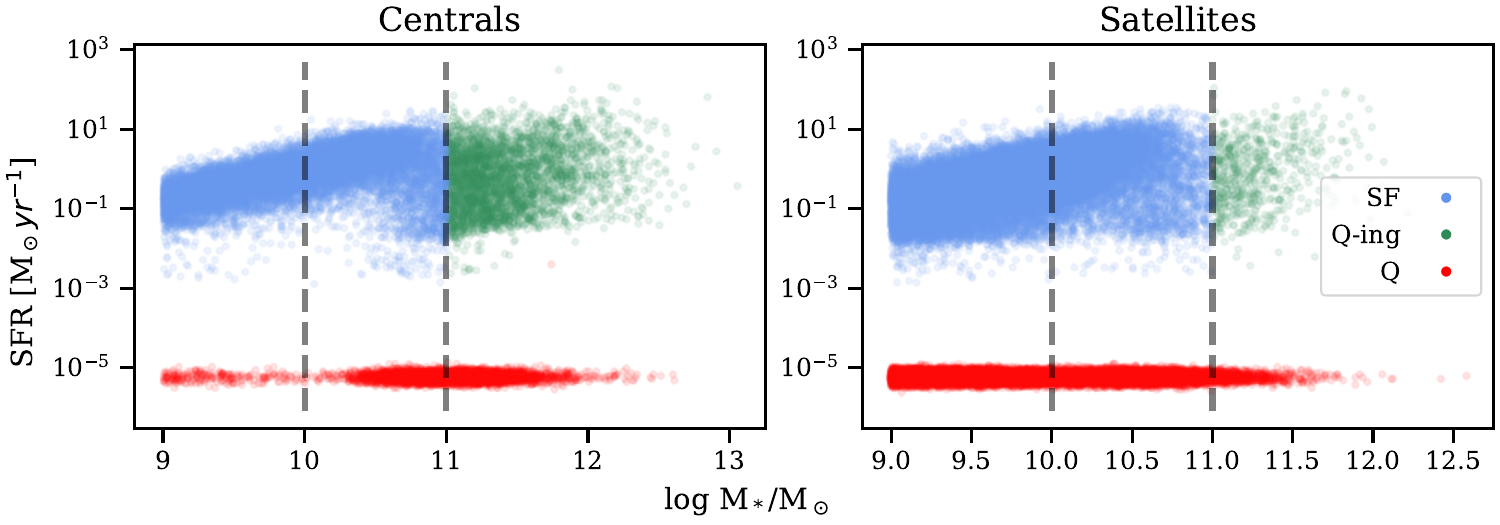}
    \caption{The star-formation rate vs stellar mass of centrals (left) and satellite (right) galaxies in the training sample. Quiescent galaxies have zero SFR in the most recent bin, spanning 475 Myrs, but they are plotted here with a dummy value for presentation values. The colours demonstrate how galaxies are classified in this diagram: quiescent, star forming, or high-mass star-forming. Also represented are the three stellar mass bins used, defined as low mass ($\log$ M$_*$/M$_\odot < 10$), intermediate mass ($10 \leq \log$ M$_*$/M$_\odot < 11$), and high mass ($\log$ M$_*$/M$_\odot \ge 11 $). }
    \label{fig:MS_gal_samples}
\end{figure*}

A purely stochastic correction would assume random $\theta_\mathrm{true}$ for each frequency of the star-formation and chemical enrichment histories. However, given the relationship between halo mass assembly and star formation, for example, we expect that the phases of these two functions are correlated to some extent, even considering a time-delay (which would translate onto a shift in phase between the two). In this section, we investigate the correlations between phases of the MAH, SFH and ZH of different populations of galaxies in the training sample with the goal of understanding the appropriate limits of our stochastic corrections. We define three populations of galaxies according to their position in the star-formation main sequence, as shown in Fig.~\ref{fig:MS_gal_samples}. The \red{star formation rate} (SFR) is computed by averaging the stellar mass in last 1 Gyr. We split the galaxy population into star-forming and quiescent, and further split off the high-mass end of the star-forming main sequence - these galaxies sit below the main sequence, and we will refer to them as "quiescing", as they are potentially quenching now. We also define a distinct set of three populations according only to their stellar mass, the boundaries of which are also shown in Fig.~\ref{fig:MS_gal_samples} and are the same boundaries used in Fig.~\ref{fig:FTphases_cen_SFH}.

\begin{figure*}
    \includegraphics[width = \textwidth]{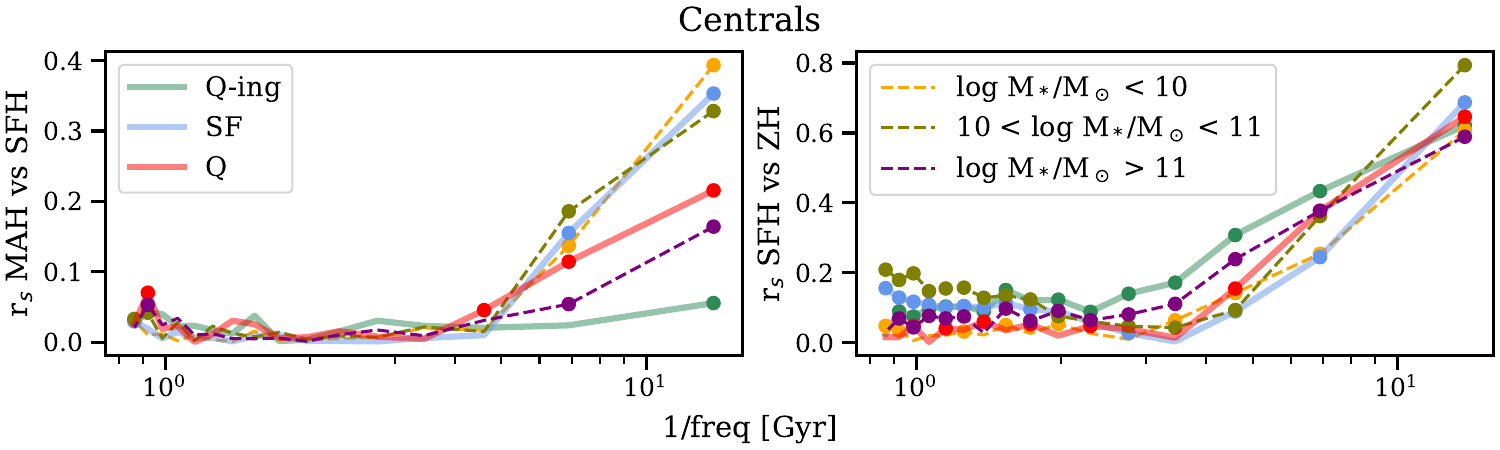}
    \includegraphics[width = \textwidth]{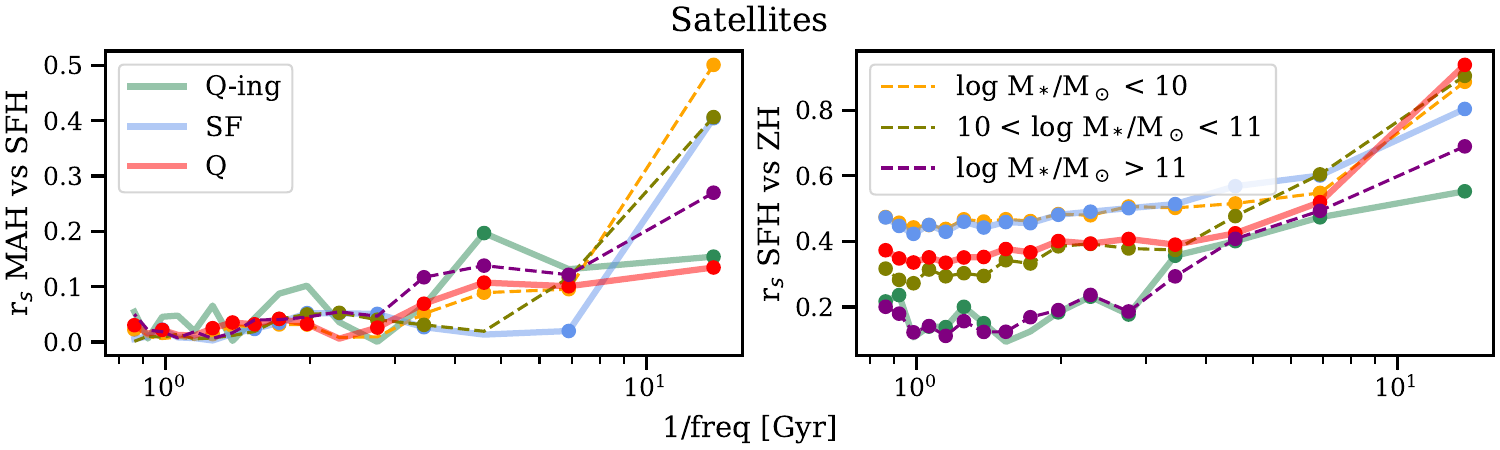}
    \caption{The Spearman rank correlation coefficient, $r_s$, between the phases of the halo mass assembly histories and the star formation histories (left) and between the phases of the star formation and chemical enrichment histories (right) as a function of frequency. In the case of satellites (bottom), we considered the subhalo mass assembly histories. Solid lines show $r_s$ of different types of galaxies, as shown in Fig.~\ref{fig:MS_gal_samples}. Dashed lines show $r_s$ of galaxies of different stellar mass. Statistically significant correlations are identified by a solid symbol. In general we find that halo and stellar growth are most correlated on long timescales ($\gtrsim 6$ Gyr) and for low-to-intermediate stellar mass galaxies in the star-forming main sequence. We also find that star-formation and chemical enrichment are correlated to shorter timescales, with less dependence on galaxy type or stellas mass. }
    \label{fig:rs_cen_sat}
\end{figure*}

To investigate to what extent the phases of the different histories are correlated, we compute the Spearman Rank correlation coefficient, $r_s$, between MAH and SFH, SFH and ZH, and MAH and ZH, as a function of frequency, for each galaxy sample and separately for centrals and satellites (Fig.~\ref{fig:rs_cen_sat}). In almost all cases, we see that the phases of the MAH and SFHs are correlated only on timescales longer than at least 6 Gyr. This likely contributes to the fact that the rNN only captures the nature of SFHs on long timescales - the rNN learns the connection between halo and stellar assembly, but Fig.~\ref{fig:rs_cen_sat} shows that these are uncorrelated on timescales shorter than 6 Gyr. More interestingly, Fig.~\ref{fig:rs_cen_sat} also reveals a dependency of the galaxy-halo connection on the type of galaxy. Specifically, galaxies in the lower-mass end of star-forming main sequence have the strongest correlation between the phases of their halo and stellar assemblies. These galaxies, sitting in the linear part of the star-forming main sequence, are often described as gas-regulated systems \citep{2013ApJ...772..119L, 2014ApJ...793...12B}, whereby galaxies are forming stars from infalling cold gas at a fixed efficiency. The expected coincidence of infalling cold gas and dark matter onto the halo then gives a plausible explanation as to why the phases of these galaxies' MAH and SFHs show the strongest correlation. At the high mass end of the star-forming sequence, however, the correlation between halo mass assembly and star formation decreases. $r_s$ is very small at {\em all} timescales for central SF massive galaxies, and effectively zero is we select galaxies based on a SFR averaged over the last 475 Myr, instead of 1 Gyr.  In these galaxies, star formation timescales have decoupled from those of mass assembly even on very long timescales. In the case of satellites, we correlate the phases of the star-formation histories with the subhalo assembly histories. The main difference is in the quenched galaxies, although we note a more complex relationship between halo and star-formation phases. Obviously our galaxy samples, defined according to either stellar mass or position in the star-forming main sequence, are not independent. The purpose of this paper is not to carefully disentangle the main drivers of the correlation between MAH, SFH and ZH phases, but rather to understand to what extent and in what conditions they correlate so that we may inform how we determine $\theta_\mathrm{true}$. Also clear from Fig.~\ref{fig:rs_cen_sat} is that the phases of star formation and chemical enrichment are substantially more strongly correlated, for all galaxy types and in particular satellite galaxies. \\

\subsection{Implementation} \label{sec:implementation}

\red{Given the results from the previous section, we decide firstly not to modify the phases of the two longest frequency components as they are well predicted by the network\footnote{\red{Fig.~\ref{fig:FTphases_cen_SFH} shows that at a timescale of 6.90Gyr, this statement is no longer true for our lowest mass bin. However, drawing stochastic phases for these galaxies at this frequency makes no significant difference to our results, and we therefore adopt the scheme here for simplicity.}}. We also preserve the amplitudes of the longest component, but only for galaxies with $\log M_*/M_\odot < 9.5$, where predictions remain accurate (see Fig.~\ref{fig:FTamp_cen}).
%\footnote{\red{This exception is motivated by the improved amplitude agreement at long timescales for low-mass galaxies, as seen in Fig.~\ref{fig:FTamp_cen}.}}. 
Secondly we decide to treat the $\theta_\mathrm{true}$ for the SFH as effectively stochastic (i.e., with no imposed correlation with the phases of the halo's accretion history) for all other frequencies. }

That leaves two important matters to consider: 1) the fact that SFH and ZH phases are effectively not random in the training sample (i.e., they do not uniformly populate the $[0,2\pi]$ interval); 2) the fact that the phases of SFH and ZHs are strongly correlated. 

We address the first by randomly sampling phases from the training sample from galaxies matched in stellar mass, therefore insuring that - although stochastic - the phases are drawn from the same range spanned by the data\footnote{\red{We tried other matching schemes between test and training data, such as a 2D match on halo and stellar mass, or a halo mass alone scheme, but none provided significant improvement on a match on stellar mass alone.}}. To achieve this, the predicted sample is first divided into 15 bins based on the percentile distribution of its stellar masses. The training sample is then binned according to the same stellar mass bins, and for each bin, a single galaxy is randomly selected from the corresponding training sample bin. This approach ensures that the selected training data spans the same range of stellar masses as the predicted sample while maintaining the stochastic nature of the phase selection. We address the second by drawing phases for SFH and ZH from the same galaxy, thereby maintaining any existing correlations between the two quantities in the training data. The subset of training sample picked this way is termed as $h_\text{subtrain}(t)$ with the corresponding FT as $H_\text{subtrain}(f)$, FT phases as $\theta_\text{subtrain}$ and FT amplitudes as $A_\text{subtrain}$.

Therefore, in summary, we employ the equations in Section \ref{sec:formalism} with the following substitutions for $\theta_\text{true}$ and $A_\text{true}$:

\begin{align}
     \theta_{\text{true}}(f) = 
    \begin{cases}
         \theta_{\text{pred}}(f) & \text{if $f$ = 0 or 0.03735525} \\
         \theta_{\text{subtrain}}(f) & \text{otherwise}
    \end{cases}
\end{align}

\begin{align}
    A_{\text{true}}(f) = 
    \begin{cases}
        A_{\text{pred}}(f) & \text{if $f$ = 0 and $\log M_*/M_\odot$< 9.5} \\
        _{\text{pred}}A(f) & \text{otherwise}
    \end{cases}
\end{align}

where $f$ is the frequency, \textcolor{black}{$A_{\text{pred}}(f)$ refers to the amplitudes obtained by applying DFT to the predicted time-domain histories, $h_\text{pred}(t)$. In contrast, $_{\text{pred}}{A}(f)$ denotes the amplitudes directly predicted by the rNN in Fourier space.}

This method brings substantial improvement to the SFHs and ZHs predicted by the rNN. However, although the mean of $_\text{pred}A(f)$ matches the mean of $A_\text{true}(f)$, the scatter in $_\text{pred}A(f)$ is substantially smaller. It is also the case that, in individual galaxies, $_\text{pred}A(f)$ can have occasional large residuals that, when paired with stochastic values for $ \theta_{\text{subtrain}}(f)$, can lead to fluctuations in the SFHs and ZHs that are unphysical. For example, sharp transitions and flat plateaus in the SFHs and ZHs (such as quenching) rely on phase coherence, which our method does not maintain.  We deal with these two issues - both stemming from the fact that $_\text{pred}A(f)$ is not a perfect match to $A_\text{true}(f)$ and the lack of phase coherence in our stochastic corrections - in two ways: a set of filters to avoid extreme deviations of the SFHs and ZHs from their predicted values (including negative values), and a normalisation to restore the expected standard deviation of the SFHs and ZHs. We detail these next.

\subsubsection{Filters} \label{sec:filters}

We apply four post-processing filters to the modified SFH and ZHs. In summary, these are:

\begin{enumerate}
\item All negative values resulting from the stochastic modifications are replaced with the rNN-predicted values.
\item At very early times ($t<2.8$ Gyr), modifications are restricted to a factor of two of the original predictions. 
\item Galaxies identified as quenched remain unmodified after quenching. 
\item Modifications to central galaxies at recent times ($t>10.8$ Gyr) are not allowed to fall below 35 per-cent of the rNN predictions.
\end{enumerate}

The justification for (i) is clear as negative values are unphysical. 

The reason for (ii) comes from the fact that the rNN does well at early times. The impact on colours and SEDs from this filter is very small. 

We justify (iii) because of the need to maintain the quenching of galaxies, which is well predicted by the rNN but clearly endangered by stochastic corrections which can create spurious SF events. We identify potentially quenched galaxies from their \red{specific star formation rate} (sSFR) values averaged over the last 1.008 Gyr (by requiring it to be less than 0.2 yr$^{-1}$ ). We confirm a galaxy as quenched through either the shape of their SFH (by requiring the logarithm of SFR at recent times to be at least two order of magnitudes below the logarithm of its peak value, SFR$_{peak}$) or through a fixed threshold value in sSFR which we set to 0.006 yr$^{-1}$. We then identify continuous "quenched time bins" by starting at the most recent time bin and moving towards earlier times. Once a bin has either a sSFR$>$0.006 yr$^{-1}$ or a SFR$>$SFR$_{peak}/100$, then we stop. Time bins identified as "quenched" are reverted back to the values of SFH and ZH predicted by the rNN. 
Our method to identify quenched galaxies and quenched plateaus in their SFHs was developed through trial and error by visually inspecting predicted and true SFHs of dozens of galaxies. We note that the task of identifying quenched galaxies and quenched plateaus in the training data is substantially easier as the simulation completely shuts down star-formation so one needs only to look for contiguous zero-valued bins. However, the rNN has difficulty in predicting zeros and has much shallower quenching, meaning that a straight cut in SFR or sSFR fails to identify the features we are looking for. \red{We therefore adopt this more relaxed, empirically motivated threshold that better captures recently quenched systems in predicted histories.}
%A detailed discussion and a visual comparison of this to the commonly used $10^{-11}\ \mathrm{yr}^{-1}$, is provided in Appendix~\ref{appendix:ssfr}, where we justify our chosen method and demonstrate its effectiveness.} 
We found the above to be reasonably robust to small changes in the values we quote here. This filter has the strongest impact in population colours and SEDs.

Finally, the justification for (iv) comes from the observation that the stochastic corrections can cause a subtle but systematic decrease of the SFHs at recent times. This is due to the rNN having particular difficulty at predicting the long frequency components of $_{pred} A$ in (mostly) star-forming, low-mass galaxies. Therefore this filter offers a mild improvement for low-mass centrals.

\subsubsection{Normalisation}\label{sec:normalisation}

The normalization procedure adjusts the corrected SFHs based on their stellar masses to match the standard deviation observed in the training set within specific timescale bins.

First, we partition the sample into 16 bins based on the logarithm of stellar mass of the predicted sample. The boundaries for these bins were determined using a linear spacing between the minimum and maximum logarithmic stellar mass values in the corrected dataset. 

For each bin, and at each time step, the corrected SFHs were normalized. This involved adjusting the corrected SFHs to have the same standard deviation as the training SFHs within the same bin. Specifically, for each mass bin b and each time step $t$, the normalization was computed as follows:
\begin{align}
    h_{\text{norm-mod},b,t}(t) = \left(h_{\text{mod},b,t}(t) - \mu_{\text{mod},b,t}\right)\frac{\sigma_{\text{subtrain},b,t}}{\sigma_{\text{mod},b,t}} + \mu_{\text{mod},b,t}
\end{align}
 
  where $\mu_{\text{mod},b,t}$ and $\sigma_{\text{mod},b,t}$ denote the mean and standard deviation of the modified set at time $t$ in the stellar mass bin $b$. Similarly, $\sigma_{\text{subtrain},b,t}$ denotes the standard deviation of the subset of training sample (refer to Section \ref{sec:implementation} for more details) at time $t$ in the stellar mass bin $b$. 

Any resulting negative values in the normalized SFHs/ZHs were replaced with the unnormalized modified values to ensure non-negative star formation rates. After normalization, the FT was applied to each galaxy's SFH/ZH to derive the amplitude of the SFH/ZH in the frequency domain.

The impact of this normalisation procedure is small in individual galaxies, but is important in restoring the scatter of the training FT amplitudes (see next section), which, as we have seen, is underpredicted by the rNN.

\section{Results}\label{sec:results}

\begin{figure*}
    \includegraphics[width = \textwidth]{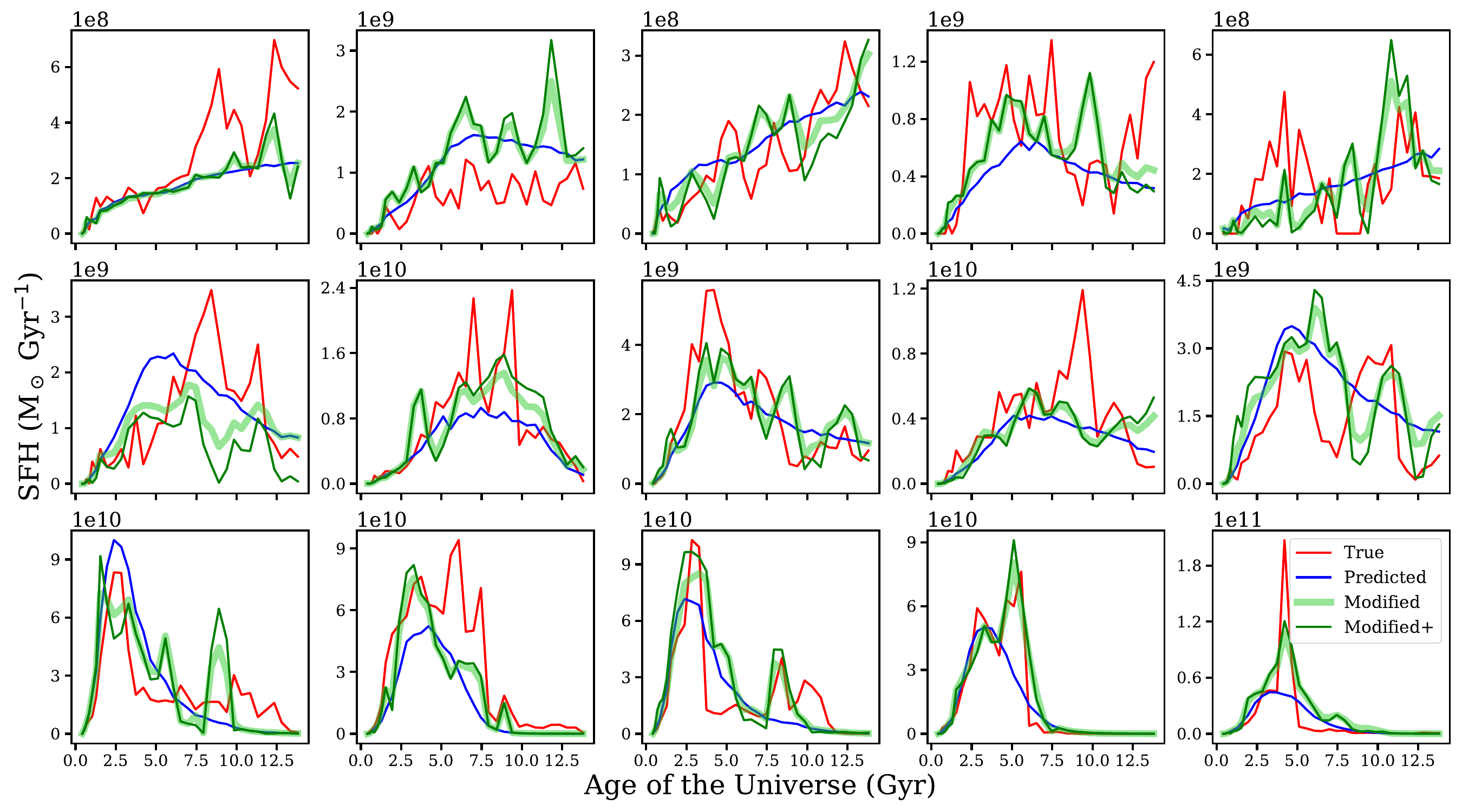}
    \caption{Star formation histories (SFHs) of 15 random central galaxies, grouped by stellar mass as a function of timescale. The top panel shows low-mass galaxies ($\log M_*/M_\odot < 10$), the middle panel shows intermediate-mass galaxies ($10 \leq \log M_*/M_\odot < 11$), and the bottom panel shows high-mass galaxies ($\log M_*/M_\odot \geq 11$). Blue lines represent the smooth rNN-predicted SFHs, red lines show the true SFHs, thick lightgreen lines show the stochastically modified SFHs and the sharp green lines show the results after normalization is applied to them. Corrections effectively restore short-timescale variability, though the specific features remain stochastic. The normalization step has a small effect on individual galaxies but ensures accurate scatter at the population level.
}
    \label{fig:cent norm sfh}
\end{figure*}

In this section we show the impact of the full set of corrections (i.e., including all filters presented in Section~\ref{sec:filters} and the normalisation step detailed in Section~\ref{sec:normalisation}) on individual SFHs, mean optical spectra, optical colours distributions, and the mass-metallicity relation. In figures we refer to results with all filters applied but no normalisation as "modified" and to results with all filters and normalisation as "modified+". In Appendix~\ref{sec:appendix-filters} we show a subset of plots that demonstrates the specific impact of the filters and normalisation.

By construction, the mean FT amplitude and standard deviation of the corrected SFHs must match that of the true data - we show this as the green line in Fig.~\ref{fig:FTamp_cen}. The stochastic corrections applied to individual SFHs can be seen in Fig.~\ref{fig:cent norm sfh} for 15 random central galaxies. The results for satellites are visually identical and not shown, and we show the same plots for ZHs in Appendix~\ref{sec:appendix-norm zh}. The blue lines show the relatively smooth SFHs predicted by the rNN and the red lines show the true SFHs. The difference between the red and blue lines demonstrates well the lack of power on small scales - even though the broad shape of the SFH is well predicted by the rNN, short timescale features are entirely absent. The SFHs after the stochastic corrections are applied (green lines) are effective at adding these features back - the specific locations of short timescale features are, of course, not reproduced with our corrections as the phases are largely stochastic. However, statistically, the impact at population level is evident as we will see later. Fig.~\ref{fig:cent norm sfh} also shows the impact of the normalisation step, detailed in Section~\ref{sec:normalisation}. On individual galaxies the effect is minor compared to that introduced by the corrections, but it guarantees that population-level quantities have the correct standard deviations.

\begin{figure*}
    \includegraphics[width = \textwidth]{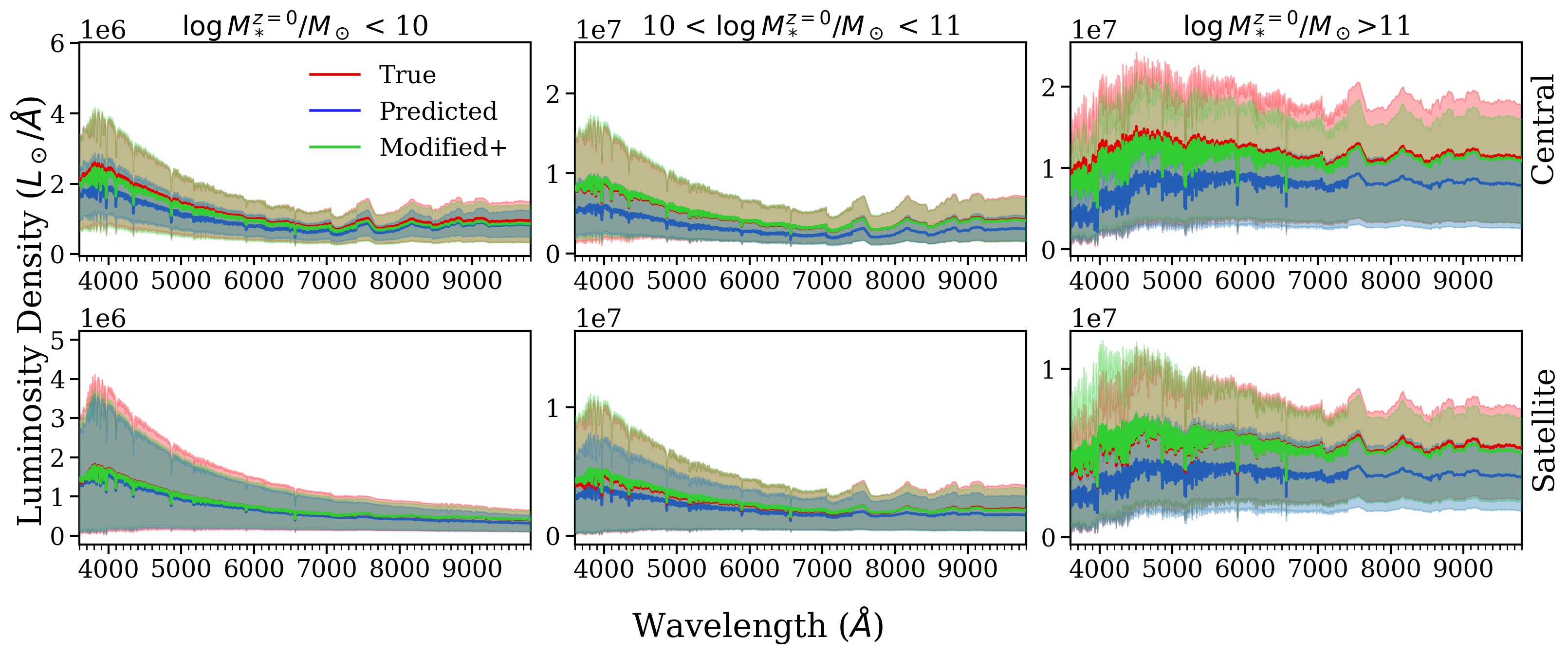}
    \caption{The \textcolor{black}{median and 15th--85th percentile range} for stacked central (top) and satellite (bottom) spectra in bins of stellar mass. Red, blue, and green represent spectra derived from the TNG, rNN-predicted, and modified star formation histories and metallicities. Emission lines are omitted for clarity. The corrections improve the overall match to the true spectra across all mass ranges, but the changes in spectral shape are subtle and better reflected in the corresponding color distributions (see Fig.~\ref{fig:norm color}).}

    \label{fig:norm spectra}
\end{figure*}

\begin{figure*}
    \includegraphics[width = \textwidth]{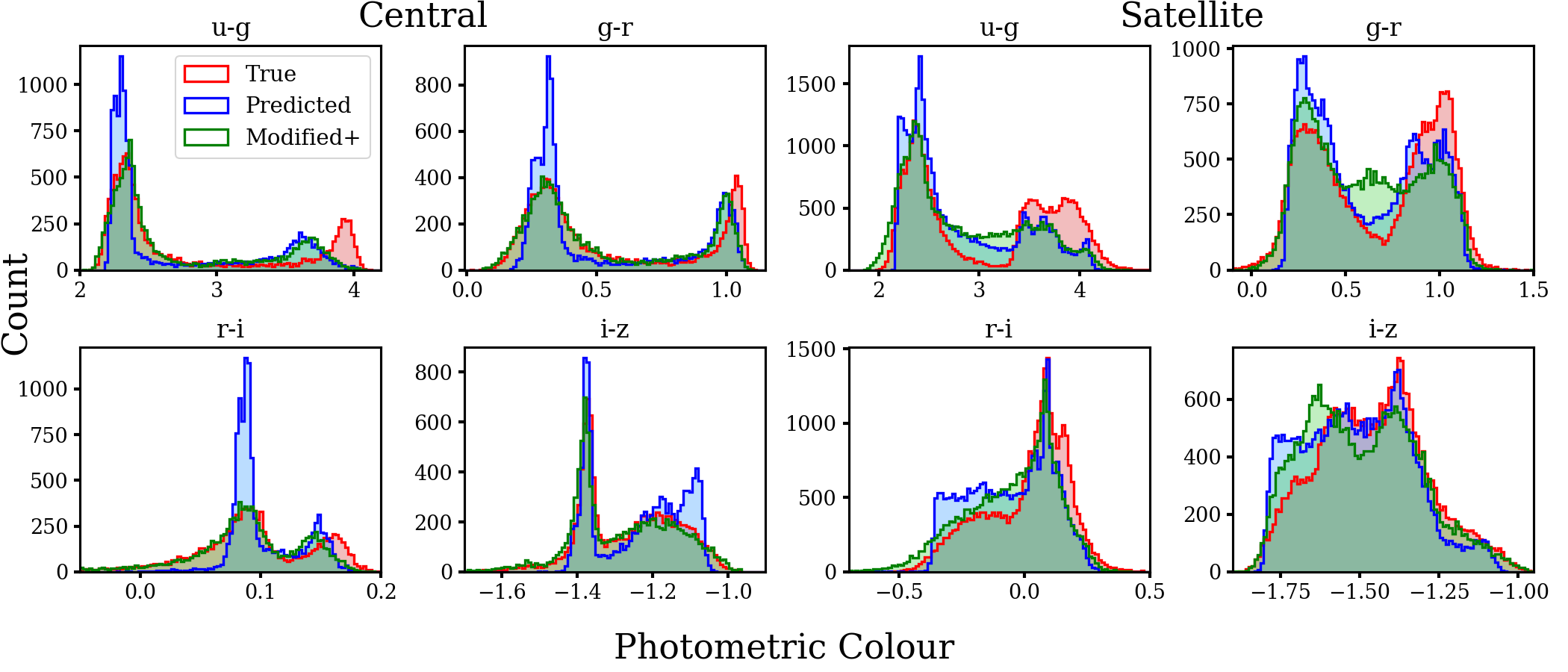}
    \caption{Photometric color distributions across five bands, showing differences between two consecutive bands. The bimodal distributions derived from TNG data are in red, rNN predictions in blue and modified spectral energy distributions in green. The corrections help recover the true color distribution for central galaxies, especially for the blue star-forming galaxies. However, in the case of satellite galaxies, the modifications slightly blur the galaxy bimodality especially in the g-r color.
}
    \label{fig:norm color}
\end{figure*}

The optical spectra and mean colour distributions are shown in Figs.~\ref{fig:norm spectra} and~\ref{fig:norm color}, respectively. Optical spectra and colours are computed according to the same methodology of CT23 - in summary we use the Flexible Stellar Population Synthesis package \citep{2009ApJ...699..486C, 2010ApJ...712..833C,2010ascl.soft10043C} to produce absorption and emission spectra according to the SFHs and ZHs in each case. For each galaxy, the optical spectra is then integrated over $g-$, $r-$, $i-$ and $z-$band filter response curves to produce rest-frame magnitude and colours. Emission lines are not shown in Fig.~\ref{fig:norm spectra} for clarity, but are included in the computation of the colours shown in Fig.~\ref{fig:norm color}. Fig.~\ref{fig:norm spectra} shows that mean and standard deviation of the optical spectra are improved after the stochastic corrections are applied (green line and shaded regions), at all mass ranges and for central and satellite galaxies. Changes in the shape of the spectra are hard to appreciate in this figure, but their impact can be seen in the colour distributions of Fig.~\ref{fig:norm color}. Here the impact of the corrections is shown to be different for different colours and also for centrals and satellites. In the case of central galaxies, the stochastic corrections help recover the true colour distribution towards the red end, but fail to move the red sequence in $u-g$ to its true position. This seems to be associated with fast-quenching galaxies, whose SFHs are hard to predict by the rNN and hard to fix with stochastic corrections - a sharp transition to a flat (quenched) plateau requires coherent combination of phases, which our method cannot accommodate. In every other case for central galaxies, the stochastic corrections bring the colour distributions in good agreement with those of the true sample. For satellite galaxies, the situation is not as clear. Improvement is more marginal and, in $g-r$, the modifications blur the galaxy bimodality in satellites. The reason seems to be related specifically with low mass satellites and the difficulty in predicting reliable FT amplitudes in that regime \citep{2024arXiv240916079C}.

\begin{figure*}
   \includegraphics[width = \textwidth]{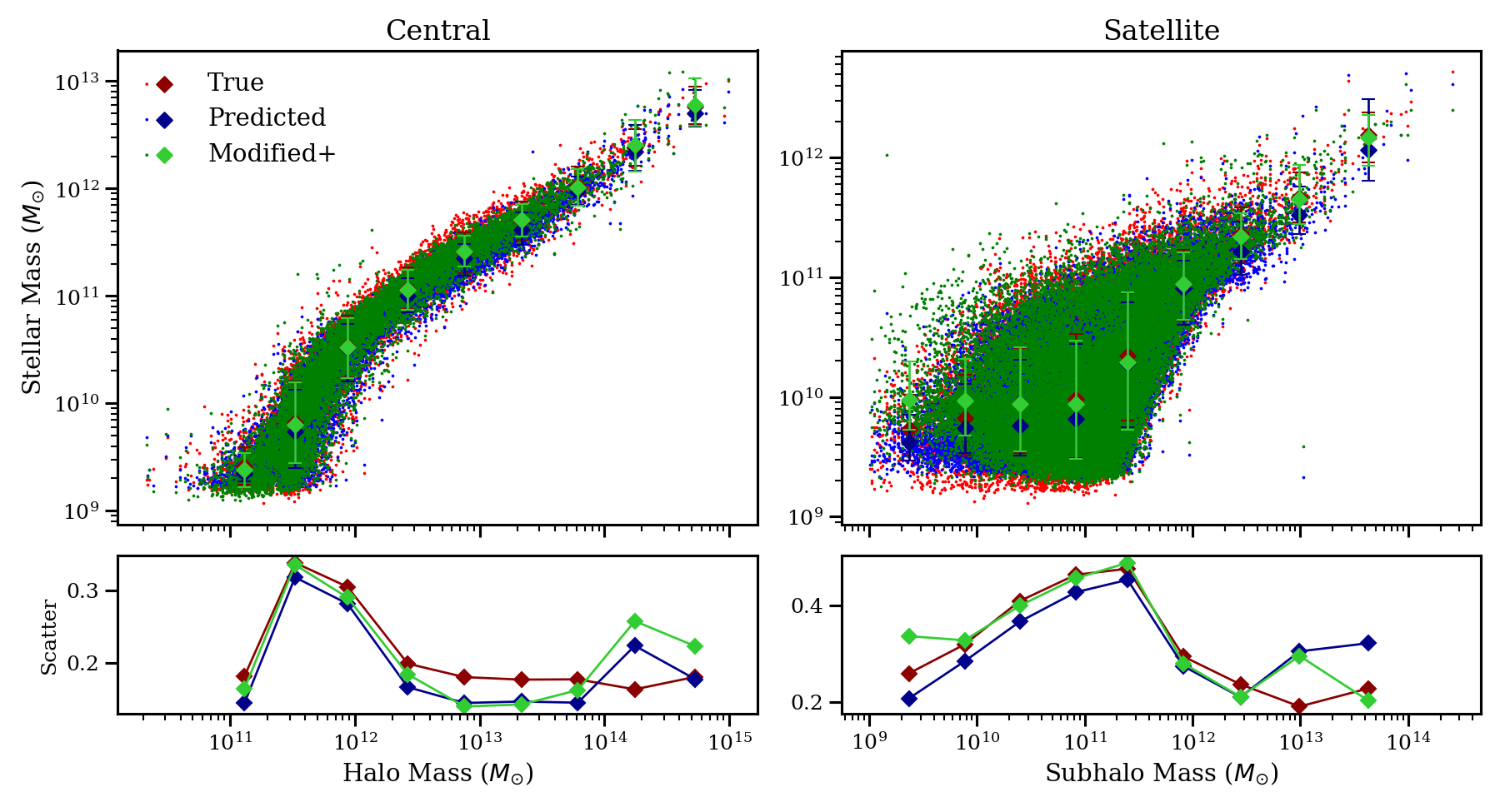}
   \caption{The top panels show the stellar-halo mass relation evaluated using the true, predicted, and stochastically modified star formation rates for central galaxies (left) and satellite galaxies (right). Stellar mass is shown as a function of FoF halo mass for centrals and subhalo mass for satellites. Data points from the original TNG data are shown in red, the uncorrected predictions in blue, and the post-normalization modifications in green. Overlaid symbols represent the binned mean stellar mass in each halo mass bin, with error bars indicating the 15th to 85th percentile range. 
   \red{The bottom panels show the scatter in stellar mass at fixed halo mass, computed as the standard deviation in log space for each bin.} The stochastic modifications bring improvement to the mean and scatter of the SHMR on all mass ranges for centrals and for satellites - see Tables \ref{tab:shmr_central} and \ref{tab:shmr_satellite}. }
   \label{fig:norm shmr}
\end{figure*}

\begin{table}
\centering
\caption{\textcolor{black}{The mean (top three rows) and scatter (bottom three rows) of stellar mass in halo mass bins for true, predicted, and modified+ data for centrals galaxies. The modifications bring a small improvement to the mean and scatter (which were already well predicted by the rNN) in all halo mass bins.}}
\label{tab:shmr_central}
\begin{tabular}{lccc}
\toprule
\textbf{Halo Mass} & \textbf{True} & \textbf{Pred} & \textbf{Mod+} \\
\hline\hline
& & \textbf{$\langle \log(M_*/M_\odot)\rangle$} & \\ 
\midrule
$\log(M_h/M_\odot)<12$        & 9.93 & 9.87 & 9.91 \\
$12 < \log(M_h/M_\odot) < 13$ & 11.07 & 11.01 & 11.07 \\
$\log(M_h/M_\odot)>13$        & 11.78 & 11.71 & 11.78 \\
\midrule
 &  &   $\sigma_{log (M_*/M_\odot)} $ & \\
\midrule
% Very Low Mass ($<$11) & 0.17 & 0.13 & 0.16 \\
$\log(M_h/M_\odot)<12$                      & 0.45 & 0.43 & 0.45 \\
$12 < \log(M_h/M_\odot) < 13$                      & 0.30 & 0.28 & 0.29 \\
$\log(M_h/M_\odot)>13$                     & 0.30 & 0.30 & 0.29 \\
\bottomrule
\end{tabular}
\end{table}

\begin{table}
\centering
\caption{\textcolor{black}{Same as Table \ref{tab:shmr_central} but for satellite galaxies. The original predictions for satellites are marginally poorer than those for centrals, leading to larger improvements from the modifications.}}
\label{tab:shmr_satellite}
\begin{tabular}{lccc}
\toprule
\textbf{SubHalo Mass} & \textbf{True} & \textbf{Pred} & \textbf{Mod+} \\
\hline\hline
& & \textbf{$\langle\log( M_*/M_\odot)\rangle$} & \\ 
\midrule
% Very Low Mass ($<$9.5) & 9.73 & 9.65 & 9.97 \\
% $\log(m_h/M_\odot)<11$        & 9.93 & 9.87 & 9.94  \\
% $11 < \log(m_h/M_\odot) < 12$ & 10.33 & 10.26 & 10.28 \\
% $\log(m_h/M_\odot)>12$        & 11.27 & 11.19 & 11.27 \\
$\log(m_h/M_\odot)<11$        & 9.93 & 9.87 & 9.94  \\
$11 < \log(m_h/M_\odot) < 12$ & 10.33 & 10.26 & 10.28 \\
$\log(m_h/M_\odot)>12$        & 11.27 & 11.19 & 11.27 \\
\midrule
 &  &   $\sigma_{log (M_*/M_\odot)} $ & \\
\midrule
% Very Low Mass ($<$9.5) & 0.25 & 0.20 & 0.34 \\
$\log(m_h/M_\odot)<11$                      & 0.42 & 0.38 & 0.41 \\
$11 < \log(m_h/M_\odot) < 12$               & 0.52 & 0.50 & 0.53 \\
$\log(m_h/M_\odot)>12$                    & 0.32 & 0.29 & 0.30 \\
\bottomrule
\end{tabular}
\end{table}

\begin{figure*}
    \includegraphics[width = \textwidth]{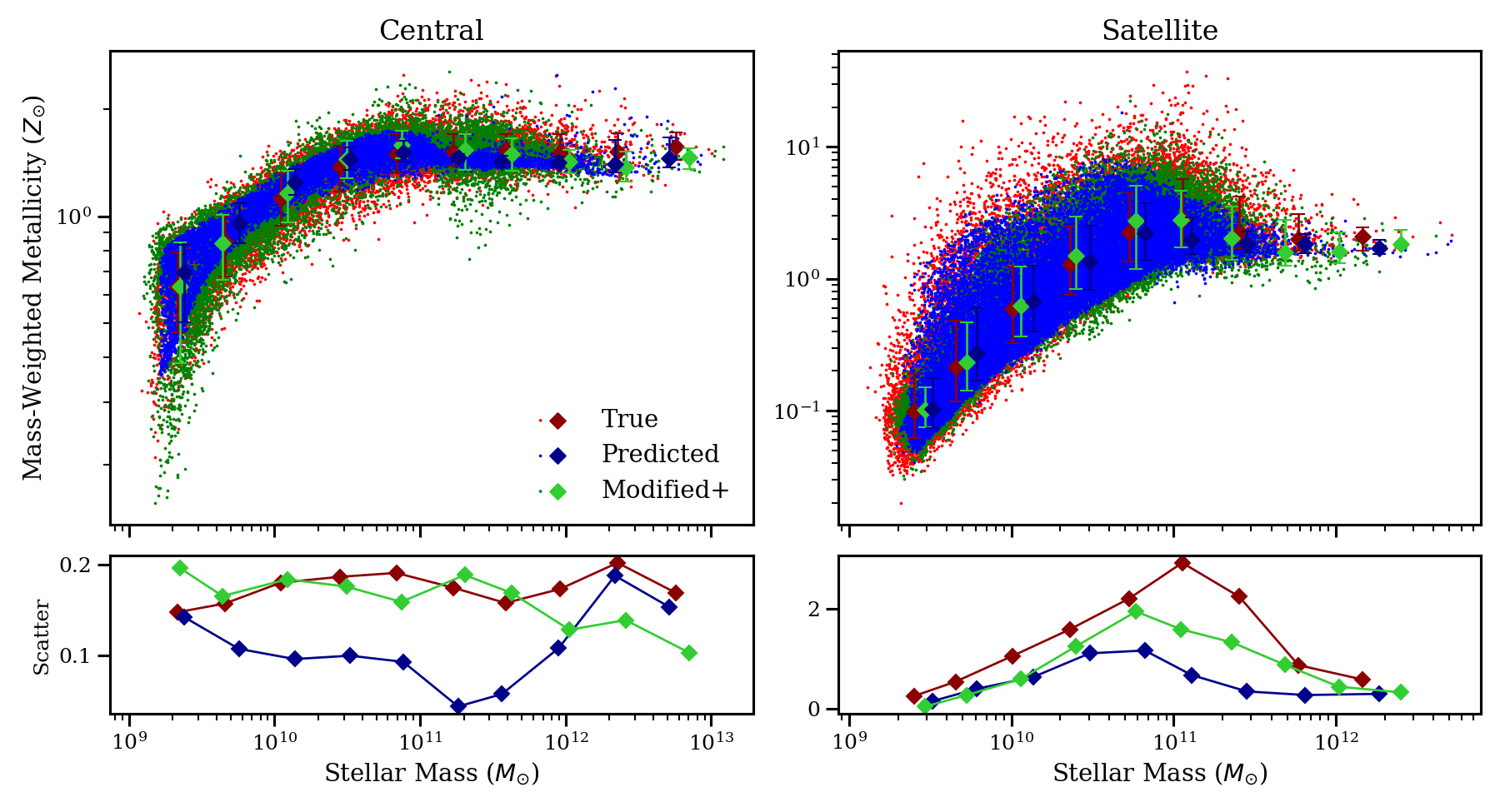}
    \caption{The top panels show the mass-metallicity relation evaluated using the true, predicted, and stochastically modified chemical enrichment histories, for central galaxies (left) and satellite galaxies (right) and shown as a function of stellar mass. Data points from the original TNG data are shown in red, predictions in blue, and the stochastically modified results in green. The overlaid markers represent the binned mean stellar metallicity, and error bars denote the 15th to 85th percentile range within each stellar mass bin. \red{The bottom panels show the scatter in metallicity at fixed stellar mass, computed as the standard deviation in log space within each bin}. While the predicted MZR captures the overall trend, it underestimates the scatter, particularly at intermediate and high stellar masses. The stochastic modifications improve the scatter for both centrals and satellites, though the full extent of variability still remains slightly underrepresented in satellites at lower masses - see Tables \ref{tab:mzr_central} and \ref{tab:mzr_satellite}.
}
    \label{fig:norm mzr}
\end{figure*}

\begin{table}
\centering
\caption{\textcolor{black}{The mean (top three rows) and scatter (bottom three rows) of the stellar metallicity of central galaxies, in bins of stellar mass, for the true, predicted and modified+ datasets. The predictions significantly underestimate the scatter of the MZR  at intermediate and high stellar masses (by around 35\% to 60\% respectively), but the stochastic modifications are able to restore the scatter at these masses.}}
\label{tab:mzr_central}
\begin{tabular}{lccc}
\toprule
\textbf{Stellar Mass} & \textbf{True} & \textbf{Pred} & \textbf{Mod+} \\
\hline\hline
& & \textbf{$\langle Z/Z_\odot \rangle$} & \\ 
\midrule
% Low Mass (9.5--10)    & 0.91 & 0.94 & 0.88 \\
$\log(M_*/M_\odot)<10$         & 0.81 & 0.82 & 0.78 \\
$10 < \log(M_*/M_\odot) < 11$  & 1.38 & 1.41 & 1.42 \\
$\log(M_*/M_\odot)>11$         & 1.55 & 1.46 & 1.52 \\
\midrule
& & $\sigma_{Z/Z_\odot}$ & \\
\midrule
% Low Mass (9.5--10)    & 0.18 & 0.13 & 0.18 \\
$\log(M_*/M_\odot)<10$         & 0.21 & 0.20 & 0.23 \\
$10 < \log(M_*/M_\odot) < 11$  & 0.22 & 0.14 & 0.23 \\
$\log(M_*/M_\odot)>11$         & 0.17 & 0.07 & 0.18 \\
\bottomrule
\end{tabular}
\end{table}

\begin{table}
\centering
\caption{\textcolor{black}{As Table \ref{tab:mzr_central} but for satellite galaxies. The predictions underestimate the mean and scatter at all masses (between around 20\% to 75\%, depending on the mass). The modifications bring substantial improvements in all cases except to the scatter of low mass satellites. Unlike the case with centrals, the stochastic corrections are not able to completely restore the mean and scatter of the true MZR.}}
\label{tab:mzr_satellite}
\begin{tabular}{lccc}
\toprule
\textbf{Stellar Mass} & \textbf{True} & \textbf{Pred} & \textbf{Mod+} \\
\hline\hline
& & \textbf{$\langle Z/Z_\odot \rangle$} & \\ 
\midrule
$9.5 < \log(M_*/M_\odot)<10$         & 0.47 & 0.33 & 0.35 \\
$10 < \log(M_*/M_\odot) < 11$  & 2.05 & 1.61 & 2.01 \\
$\log(M_*/M_\odot)>11$         & 3.40 & 2.09 & 2.81 \\
\midrule
& & $\sigma_{Z/Z_\odot}$ & \\
\midrule
$9.5 < \log(M_*/M_\odot)<10$         & 0.69 & 0.37 & 0.34 \\
$10 < \log(M_*/M_\odot) < 11$  & 1.99 & 1.18 & 1.66 \\
$\log(M_*/M_\odot)>11$         & 2.74 & 0.62 & 1.46 \\
\bottomrule
\end{tabular}
\end{table}

Fig.~\ref{fig:norm shmr} shows the stellar-to-halo relation (SHMR) for centrals and satellites. \red{Tables \ref{tab:shmr_central} and \ref{tab:shmr_satellite} show the mean stellar mass and scatter, in bins of halo mass, for the true, predicted and modified+ datasets. Although the original predictions were already very good (within 0.1 dex), the stochastic corrections bring further (although modest)} improvement to the mean and scatter of this fundamental relation. Finally, we consider the impact of the stochastic corrections on the mass-metallicity relation (MZR), which is shown in Fig.~\ref{fig:norm mzr}. One of the biggest limitations of the SFHs and ZHs recovered by the rNN was the difficulty in predicting the mean and scatter in the MZR. The MZR is plotted as a mass-weighted stellar metallicity as a function of stellar mass (in turn computed by integrating the SFH for each galaxy), so it depends on the accuracy of {\it both} the SFH and the ZHs. \red{Tables \ref{tab:mzr_central} and \ref{tab:mzr_satellite} show the mean metallicity and its scatter in bins of stellar mass.} The stochastic corrections bring substantial improvement \red{to the mean and to the scatter} of the MZR in centrals and satellites. \red{The scatter in the MZR, in particular, was underpredicted by around 40\% at intermediate masses (more at higher masses) in centrals and in satellites. Our stochastic corrections alleviated this issue (almost entirely in the case of centrals),} demonstrating the importance of short-time scale events in order to get the scatter of this fundamental relation correctly. \red{In the case of satellites, the modifications were not able to completely restore the true mean and scatter.}

\section{Summary and Conclusions}\label{sec:conclusions}

In this paper, we introduce an investigation of the galaxy-halo connection as a function of frequency with the intention of producing stochastic contributions to SFHs and ZHs. Our work is done in the context of the rNN predictions of CT23 that, although being successful at predicting the broad shape of SFHs and ZHs, were unable to predict short-term variability. We argued here that the cause must be, at least partly, that the phases of the mass assembly and star-formation/chemical enrichment histories are themselves decoupled on short timescales. We capitalized on the success of the rNN to directly predict the FT amplitudes of the SFHs/ZHs in order to introduce a formalism to compute semi-stochastic corrections. 

Our main results can be summarised as follows:

\begin{itemize}
\item In TNG-100, the phases of the halo mass assembly and star-formation histories are correlated only on timescales greater than around 6 Gyr (Fig.~\ref{fig:rs_cen_sat}, left). This is in agreement with a body of literature that associates short timescale contributions to star-formations histories with feedback processes or giant molecular cloud timescales, and intermediate to long timescales with gas accretion and depletion times (e.g.~\citealt{2015MNRAS.447.3548S, 2019MNRAS.487.3845C, 2020MNRAS.498..430I}). Here we attempt a more direct link between halo and star-formation histories by correlating the phases of these two components.
\item The correlation between the halo mass assembly and star-formation history phases depend on stellar mass, galaxy type, and on whether a galaxy is a central or satellite, being almost non-existent for massive quenching galaxies (Fig.~\ref{fig:rs_cen_sat}, left). The reason for the decoupling between halo and stellar growth on these galaxies, even on long timescales, is not clear. On long timescales, star-forming galaxies can be approximated as gas regulated systems, forming stars at fixed efficiency. A coherent accretion of dark matter and cold gas should then result in a correlation of the phases of halo and stellar growth. A decoupling might signify a break in star-formation efficiency or complex dependencies (for example, as a function of environment) that conspire to give uncorrelated phases on the whole. Studying this decoupling at higher temporal resolution, as a function of environment, and on different simulations will be subject of future work.
\item The phases of the star-formation and chemical-enrichment histories are correlated over a wide range of the timescales probed, with a notable dependence on stellar mass, particularly at shorter timescales (Fig.~\ref{fig:rs_cen_sat}, right). In low-mass galaxies, chemical enrichment and star-formation are coupled on all timescales, with the correlation decreasing with stellar mass. 
\item The stochastic corrections are successful at introducing power on short timescales and producing more realistic star-formation and chemical enrichment histories (Fig.~\ref{fig:cent norm sfh} and \ref{fig:cent norm zh}). 
\item The stochastic corrections improve optical spectra and colour distributions, especially for central galaxies (Fig.~\ref{fig:norm spectra} and~\ref{fig:norm color}). The stochastic corrections are unable to shift the red sequence in $u-g$ for central galaxies, which remains too blue. This is associated with fast-quenching galaxies which the rNN is unable to reproduce and the stochastic corrections (by construction) are unable to address. 
\item The improvement in the MZR afforded by the stochastic corrections (Fig.~\ref{fig:norm mzr} and \red{Tables \ref{tab:mzr_central} and \ref{tab:mzr_satellite}}) points to the importance of short timescale contributions to the scatter in this fundamental relation. Unlike the gas-phase metallicity, a mass-weighted stellar metallicity is sensitive to the full star-formation history of a galaxy, potentially making the stellar MZR a useful probe of the limits of the galaxy-halo connection. \red{In the case of satellites, the semi-stochastic corrections were not able to fully restore the true mean and scatter, which may be related to difficulties in predicting the FT amplitudes of the ZHs for satellites.}
\end{itemize}

\red{The stochastic modeling technique presented here provides a flexible way to incorporate unresolved variability into models of galaxy formation that do not physically model those scales. We demonstrate our method on the predictions of a machine-learning model, but the technique is broadly applicable to empirical and semi-analytic models (SAM) as well. 
The time-resolution of a SAM is tied to the time-steps used to compute the baryonic properties of galaxies. This resolution is dependent on the dark matter simulation, but if we take the Millennium simulation as an example \citep{2005Natur.435..629S}, the time resolution is given by 58 snapshots $\times$ 20 time steps in each, at an average time step of $10^7$ years \citep{2015MNRAS.451.2681S}. Most SAMs will compute SEDs on the fly and store only the “current” SED at a specific snapshot. Without full SFHs/ZHs, it would be difficult to implement our method. L-Galaxies \citep{2015MNRAS.451.2663H} is an exception in that it saves SFH and ZH of every galaxy, and produces SEDs in post-processing. The time resolution at which SFHs and ZHs are saved is chosen in a way that balances computational restrictions and the accuracy of galaxy magnitudes \citep{2015MNRAS.451.2681S}, using a scheme that maintains high resolution at young ages but degrades it at older stellar ages. At $z=0$, this results in 19 bins with a width between $1.5\times 10^7$ years at young ages (which matches the time step resolution) and $2.1\times 10^9$ years at old ages (which is a degradation of the time step resolution). The approach proposed in our paper could be used to add stochasticity on time-scales shorter than the timestep width, or introduce additional stochasticity on longer timescales. Our approach requires an estimate of the true amplitude of the power spectra of the SFHs/ZHs over the timescales of interest. These could be estimated directly from cosmological hydrodynamical simulations, making our method a potentially efficient way to combine information from expensive hydrodynamical simulations with fast SAM models.}

\red{The properties of {\it individual} galaxies are, in general, not significantly improved. That is expected given the semi-stochastic nature of our corrections. The key point from this study is that, on timescales shorter than around 6 Gyr, stellar and chemical enrichment histories have decoupled from halo growth. Therefore from a galaxy-halo connection perspective, it becomes irrelevant when short-timescale events happen. Provided that we have an estimate for the frequency of decoupling, we can always stochastically correct SFHs and ZHs, within the constraints shown in this paper (namely the correlation between SFH and ZH phases, and the physical range of phase values, which depends on frequency).}

\red{Understanding in detail how the phases of MAHs decorrelate from the phases of SFHs/ZHs can help GHC models produce more realistic predictions, which our method demonstrates (though there will be other ways to implement this). We see this as a first step only - Fig.~\ref{fig:rs_cen_sat} clearly shows that this decoupling happens differently for different types of galaxies. We do not know why this is yet, and we have not taken that into account in our corrections, which we leave for future work.}

Finally, our formalism for the stochastic correction can be cast as a simple empirical semi-stochastic model that parametrizes the galaxy-halo connection as a correlated time-series between halo assembly, star formation and chemical enrichment histories. Our results suggest that fundamental relations, such as the MZR, might have the sensitivity to {\it observationally constrain} the timescale at which the phases of halo assembly and star formation decouple, for example, and how that might change with galaxy type. \red{Such a measurement would be an entirely new way to test hydrodynamical models of galaxy formation.} We leave that investigation for future work. 

\section*{Acknowledgements}

We would like to thank the anonymous reviewer for their careful and detailed report, which led to several clarifications and improvements throughout the paper.

JB is grateful for support from the US Department of Energy via grants DE-SC0021165 and DE-SC0011840. JB is partially supported by the NASA ROSES grant 12-EUCLID12-0004. The UKRI Science and Technology Facilities Council supported HGC under grant ID ST/T506448/1, which the authors gratefully acknowledge. 

This research used resources of the National Energy Research Scientific Computing Center (NERSC), a U.S. Department of Energy Office of Science User Facility located at Lawrence Berkeley National Laboratory, operated under Contract No. DE-AC02-05CH11231. 

We wish to thank the IllustrisTNG project for access to their data, the extensive documentation, and their excellent user support.

This research has made use of NASA’s Astrophysics Data System and the arXiv open-access repository of electronic preprints and postprints.

JB thanks Lado Samushia for useful discussions and comments. RT acknowledges helpful conversations with Amelie Saintonge that partially influenced the direction of this paper.

%%%%%%%%%%%%%%%%%%%%%%%%%%%%%%%%%%%%%%%%%%%%%%%%%%
\section*{Data Availability}

This research uses data that is publicly available from IllustrisTNG: \url{https://www.tng-project.org/data/}. Data produced by the neural networks of CT23/CBT25 and the modifications presented in this paper are available in a Zenodo repository: \url{https://zenodo.org/records/15589166}. 



\bibliography{sources} % if your bibtex file is called example.bib
\bibliographystyle{mnras}

%\begin{thebibliography}{99}
%\end{thebibliography}

%%%%%%%%%%%%%%%%%%%%%%%%%%%%%%%%%%%%%%%%%%%%%%%%%%

%%%%%%%%%%%%%%%%% APPENDICES %%%%%%%%%%%%%%%%%%%%%

\appendix
\section{Analysis of semi-stochastic corrections}\label{sec:appendix-filters}

In this appendix, we analyze supplementary figures to illustrate the impact of applying our formalism (Section \ref{sec:formalism} and \ref{sec:phase correlations}) under different cases of implementation (Section \ref{sec:filters} and \ref{sec:normalisation}) and emphasize the importance of each filter and normalization in improving the model's ability to capture variability and scatter. We present four sets of results to justify the need for each:

\begin{itemize}
    \item Case I: Method + (i)
    \item Case II: Method +  (i) + (ii) + (iii) + (iv)
    \item Case III: Method + (i) + (ii) + (iv) + Norm
    \item Case IV: Method + (i) + (ii) + (iii) + Norm

\end{itemize}

We focus only on central galaxies (except for Case II that presents both centrals and satellites) where the effects of each filter and normalization are more pronounced. Only figures that clearly highlight differences and justify the corrections are included.

\subsection{Case I}

In this case, we apply stochastic corrections with only Filter(i), which replaces any negative values generated during the stochastic modifications on the original rNN-predicted values. This serves as a basic setup to investigate the effects of corrections without more refined filtering or normalization steps.

\begin{figure*}
    \includegraphics[width = \textwidth]{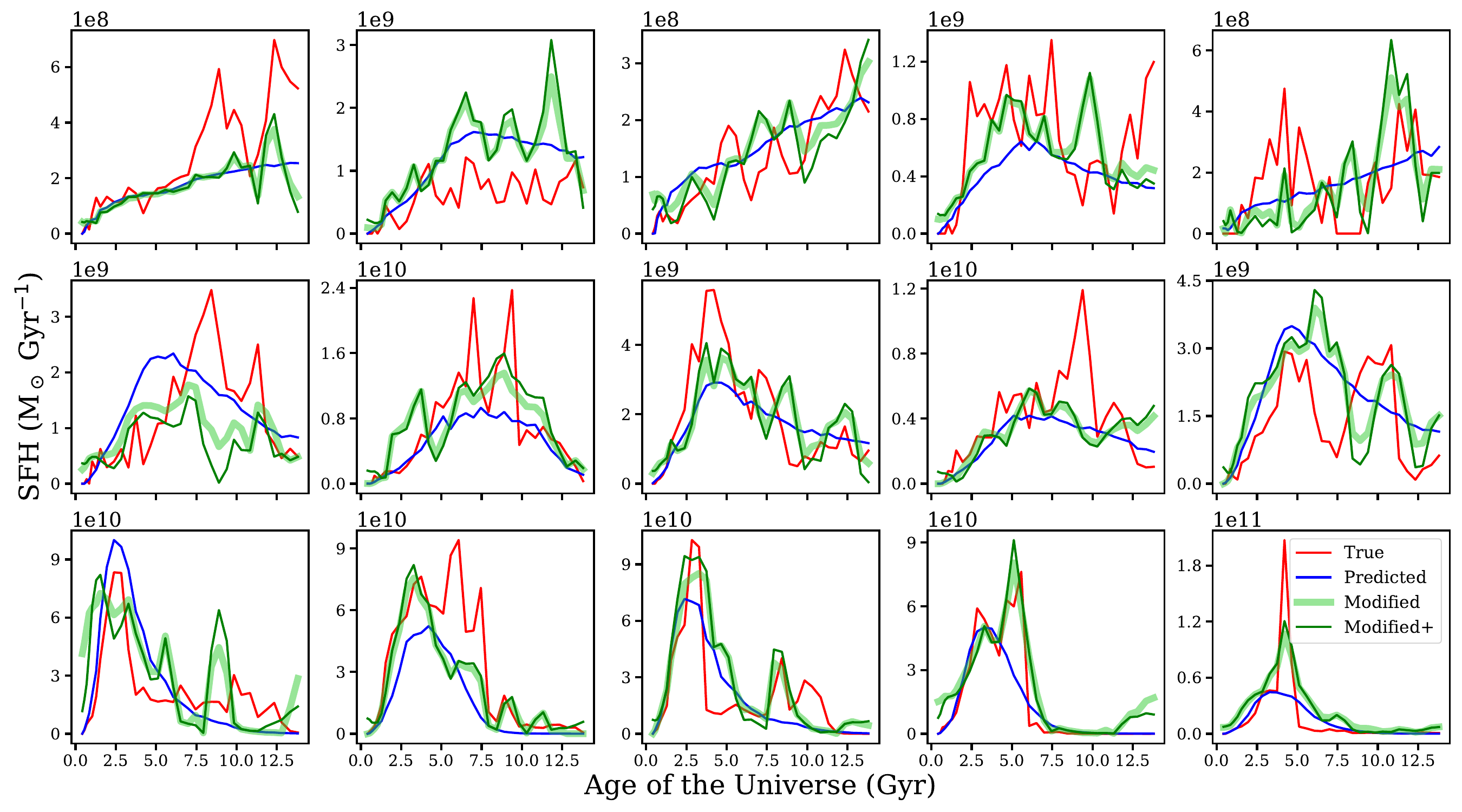}
    \caption{
Same as Figure \ref{fig:cent norm sfh}, but when the correction method excludes all filters(except filter(i)) and normalization. The top panel shows low-mass galaxies ($\log M_*/M_\odot < 10$), the middle panel shows intermediate-mass galaxies ($10 \leq \log M_*/M_\odot < 11$), and the bottom panel shows high-mass galaxies ($\log M_*/M_\odot \geq 11$). The modified SFHs start from unphysically high values, particularly at early times, due to how the Fourier transform and its inverse operate without constraints. This highlights the necessity of filter (ii) to prevent such unrealistic initial conditions.
}
    \label{fig: casei sfh}
\end{figure*}

\begin{figure*}
    \includegraphics[width = \textwidth]{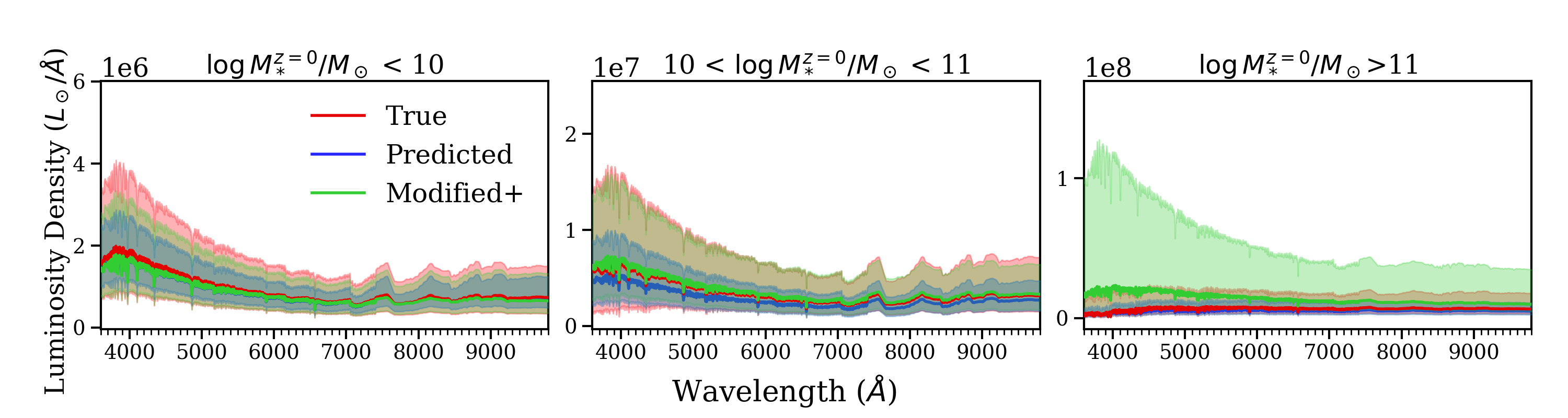}
    \caption{SEDs for central galaxies when the correction method excludes all filters(except filte(i)) and normalization. The true spectra are shown in red, the rNN-predicted spectra in blue, and the stochastically modified spectra in green. For low-mass galaxies ($\log M_*/M_\odot < 10$), there is negligible improvement as the corrections cannot significantly alter the underestimated rNN predictions. In high-mass galaxies ($\log M_*/M_\odot \geq 11$), the modifications introduce excessive variability, leading to unrealistically high spectra due to unaddressed quenching effects. Intermediate-mass galaxies show balanced spectra on average, but this balance results from offsetting over- and undercorrections, masking individual inaccuracies.
}
    \label{fig: casei spectra}
\end{figure*}

Fig. \ref{fig: casei sfh} presents the star formation histories (SFHs) for 15 random central galaxies, categorized by stellar mass similar to Fig \ref{fig:cent norm sfh}. But in this case the modified SFHs start from unphysically high values due to the limitations of the FT and inverse FT methods, which fail to address the early-time star formation features effectively. This highlights the necessity of Filter (ii), which is designed to mitigate this issue.

Fig. \ref{fig: casei spectra} shows the spectra distributions for central galaxies. For low-mass galaxies, the corrections show negligible improvement while in high-mass quenched galaxies, they introduce excessive variability leading to excessive overestimation. For intermediate-mass galaxies, the corrections seem to match the true distribution at first glance, but this is misleading. The apparent improvement in the average spectra results from balancing over- and under-corrections across different galaxies. While the mean spectra looks reasonable, the variability introduced is inconsistent, leaving individual galaxies either over- or under-corrected. Thus, without proper filtering, the overall improvement remains superficial and doesn't reflect true accuracy.

% \begin{figure}
%     \includegraphics[width = \columnwidth]{figures/Section_4/central/mod_color_magnitude_valid33.png}
%     \caption{To DO : Could be confusing to add this}
%     \label{fig: casei color}
% \end{figure}

\subsection{Case II}

\begin{figure*}
    \includegraphics[width = \textwidth]{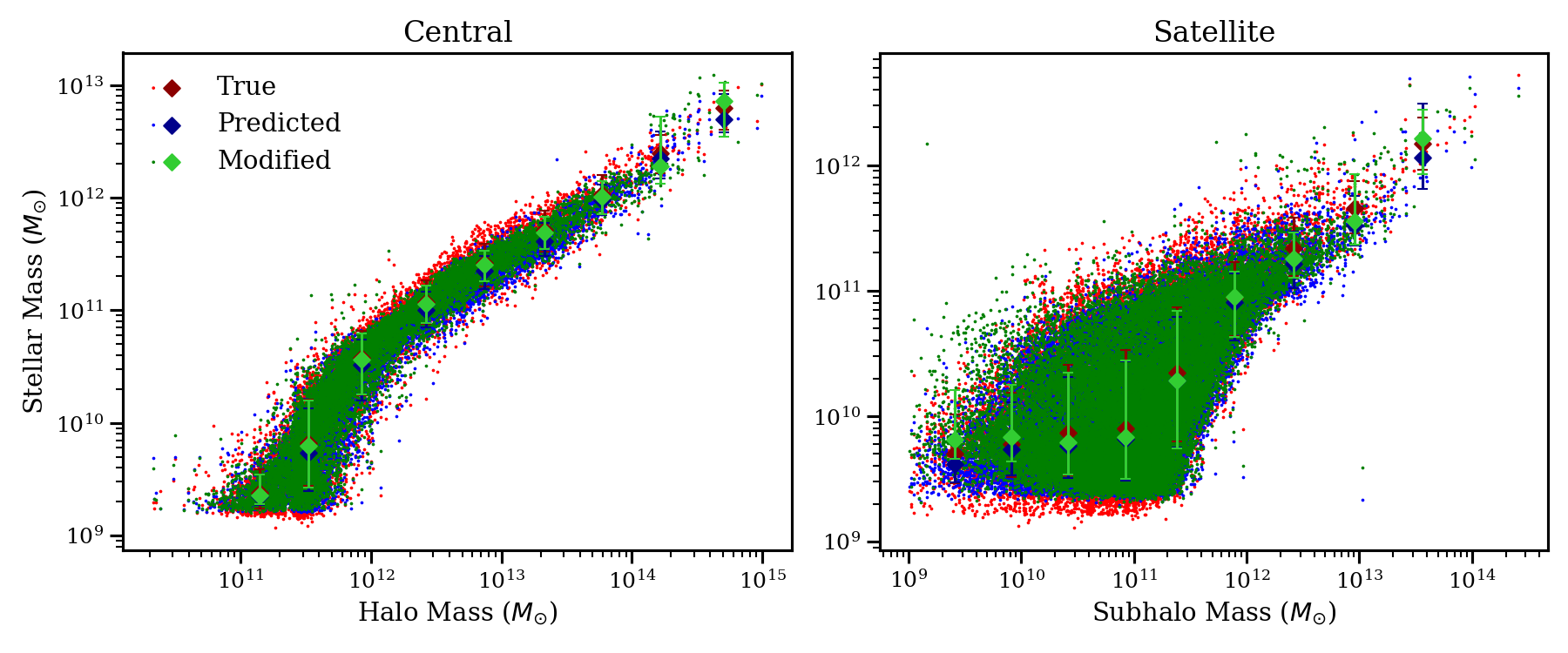}
    \caption{The stellar-halo mass relation (SHMR) for central (left) and satellite (right) galaxies similar to Fig. \ref{fig:norm shmr} but here the correction method excludes normalization. The corrections show significant improvement in the SHMR where the modifications better capture the scatter and alignment with true values, although slight discrepancies remain for especially for high-mass ($\log M_*/M_\odot \geq 11$) satellite galaxies, where variability is still not fully recovered compared to \ref{fig:norm shmr}.}
    \label{fig:mod shmr}
\end{figure*}

\begin{figure*}
    \includegraphics[width = \textwidth]{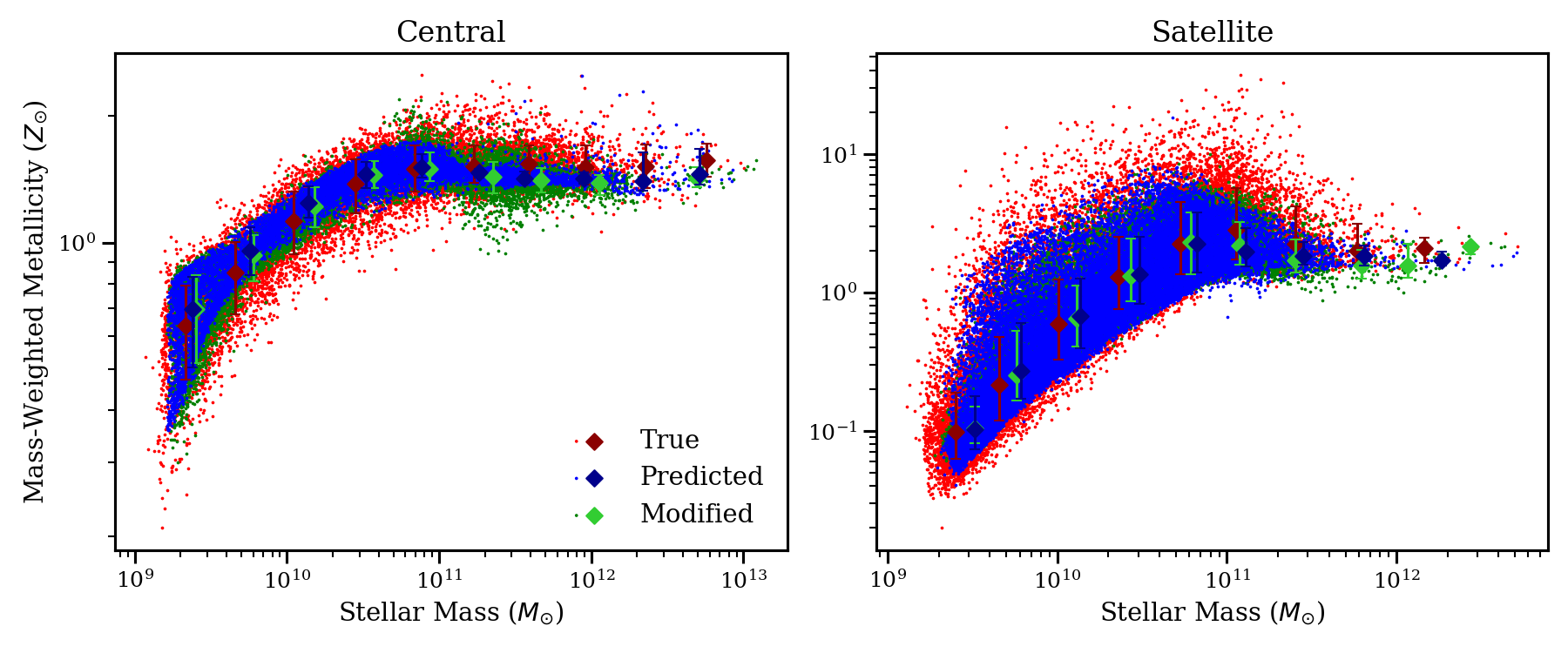}
    \caption{The mass-metallicity relation (MZR) for central (left) and satellite (right) galaxies, similar to Fig.\ref{fig:norm mzr}, but with the correction method excluding normalization. Although these corrections show significant improvement in the stellar mass (Fig.\ref{fig:mod shmr}), the improvement is not equally reflected in the metallicity. There is a notable lack of recovery in both the mean and scatter in all mass ranges, compared to the normalized results in Fig.\ref{fig:norm mzr}}
    \label{fig:mod mzr}
\end{figure*}

\begin{figure*}
    \includegraphics[width = \textwidth]{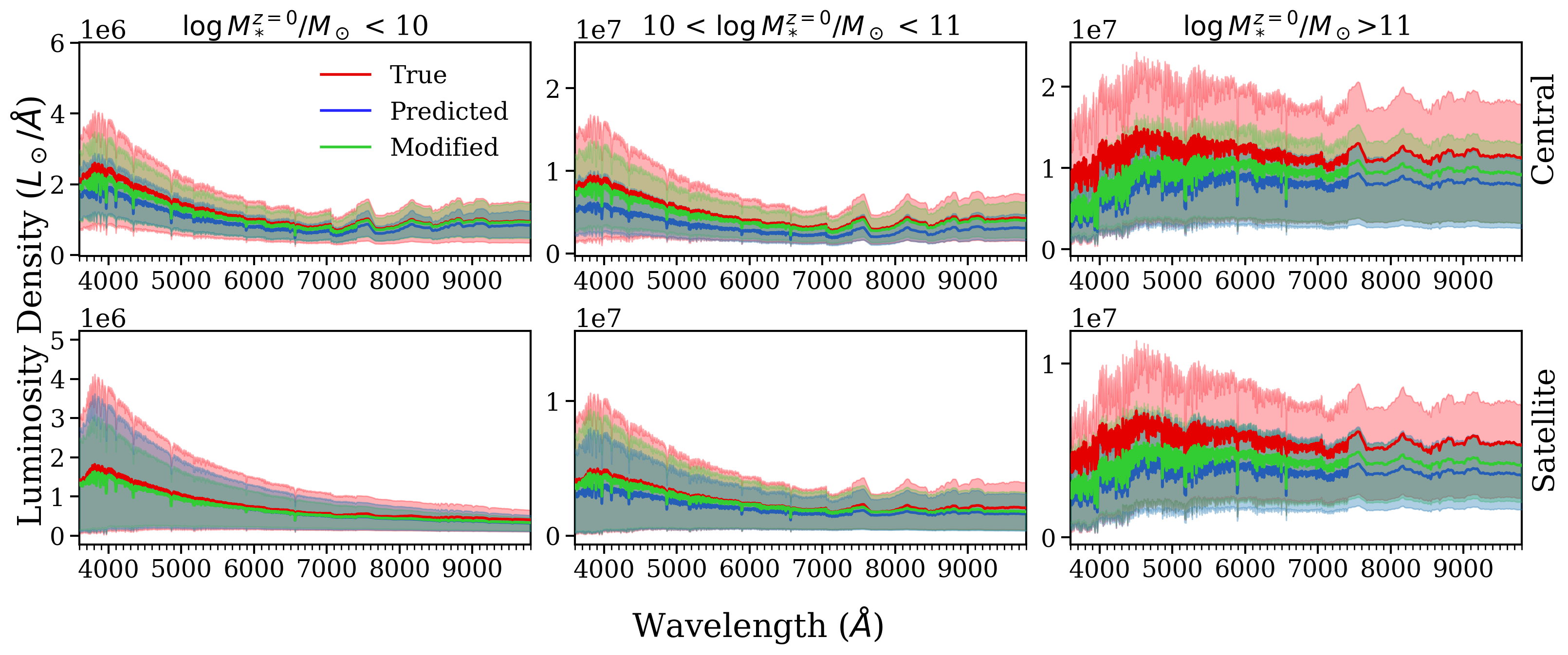}
    \caption{Spectral energy distributions (SEDs) for central (top) and satellite (bottom) galaxies, similar to Fig.\ref{fig:norm spectra}, but with the correction method excluding normalization. Although the corrections significantly improve the match to the true spectra, particularly in low-intermediate mass central galaxies and intermediate mass satellite galaxies, some discrepancies still remain in the amplitude and shape, especially for high-mass galaxies ($\log M_*/M_\odot \geq 11$). This highlights the need for normalization to properly recover both the mean and scatter in the spectra.
}
    \label{fig:mod spectra}
\end{figure*}

Fig. \ref{fig:mod shmr} shows significant improvement in the SHMR, especially in the low and intermediate mass ranges, as the model better aligns with true values by introducing necessary variability in the SFH. However, the MZR (Fig. \ref{fig:mod mzr}) remains less effective, particularly in the scatter. Similar trends are seen in the SEDs (Fig. \ref{fig:mod spectra}) where corrections improve the match to the true spectra in both mean and scatter, though some discrepancies in amplitude and shape remain. These gains are achieved through normalization (Section \ref{sec:normalisation}) that aligns the variability and scatter with true data, without altering overall trends.

\subsection{Case III}
\begin{figure*}
    \includegraphics[width = \textwidth]{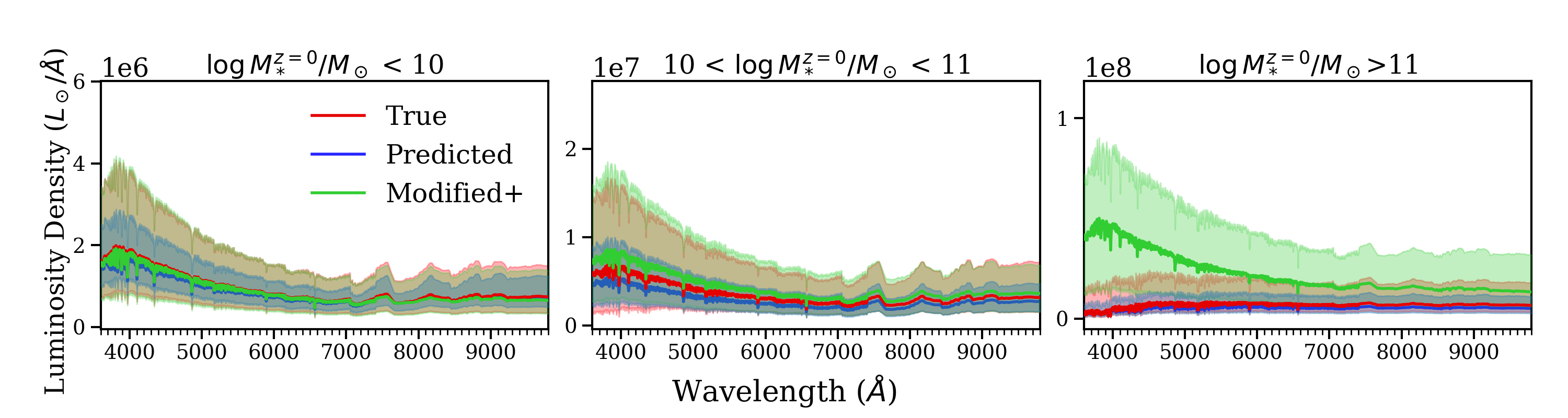}
    \caption{Spectral energy distributions (SEDs) for central galaxies when filter (iii) is excluded. The absence of this filter has minimal impact on low-mass galaxies due to the lack of quenched populations. However, for higher-mass galaxies, the correction method introduces artificial starburst events in quenched regions, leading to significantly overestimated spectra. This demonstrates the importance of filter (iii) for accurately modeling star formation in quenched galaxies.
}
    \label{fig: caseiii spectra}
\end{figure*}

\begin{figure}
    \includegraphics[width = \columnwidth]{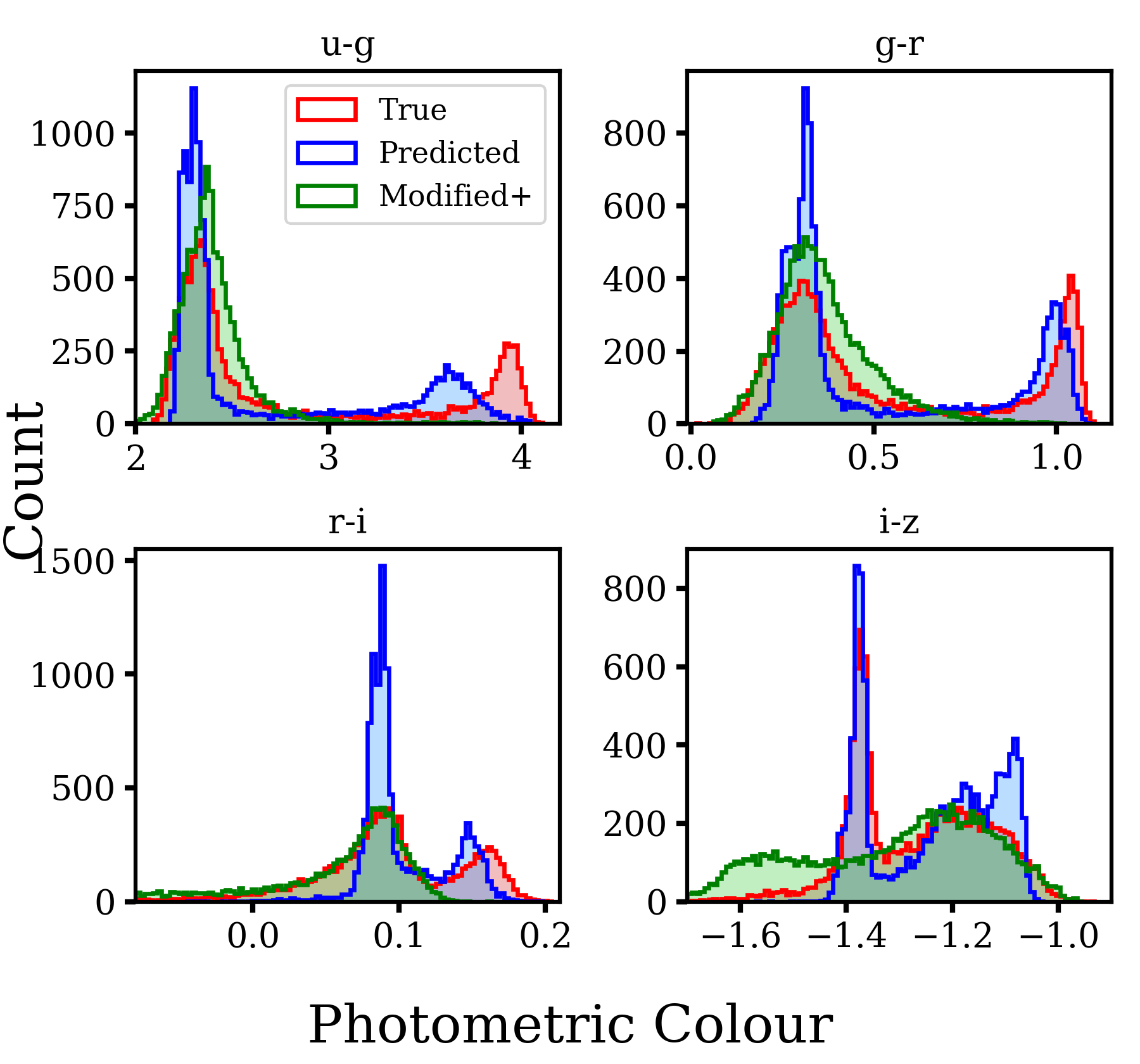}
    \caption{Color distributions for central galaxies where filter (iii) is excluded. While corrections improve the distribution for blue, star-forming galaxies, they worsen the red, quenched galaxy distribution. Without the restoration of quenched scales, artificial starburst events are introduced, resulting in incorrect colors for quenched galaxies and an increased scatter in the red color bands.
}
    \label{fig: caseiii color}
\end{figure}

Fig. \ref{fig: caseiii spectra} shows the effect of stochastic corrections on the spectra for central galaxies when filter (iii) is excluded. Comparison to Fig. \ref{fig:norm spectra} shows that this filter has negligible impact in low-mass galaxies, due to the lack of quenched galaxies. However, for higher masses, the correction formalism introduces artificial starburst events in quenched regions, leading to significantly overestimated spectra. This is further reflected in Fig. \ref{fig: caseiii color}, where this implementation improve the color distribution of blue, star-forming galaxies but worsen the red, quenched galaxies.

\subsection{Case IV}

\begin{figure*}
    \includegraphics[width = \textwidth]{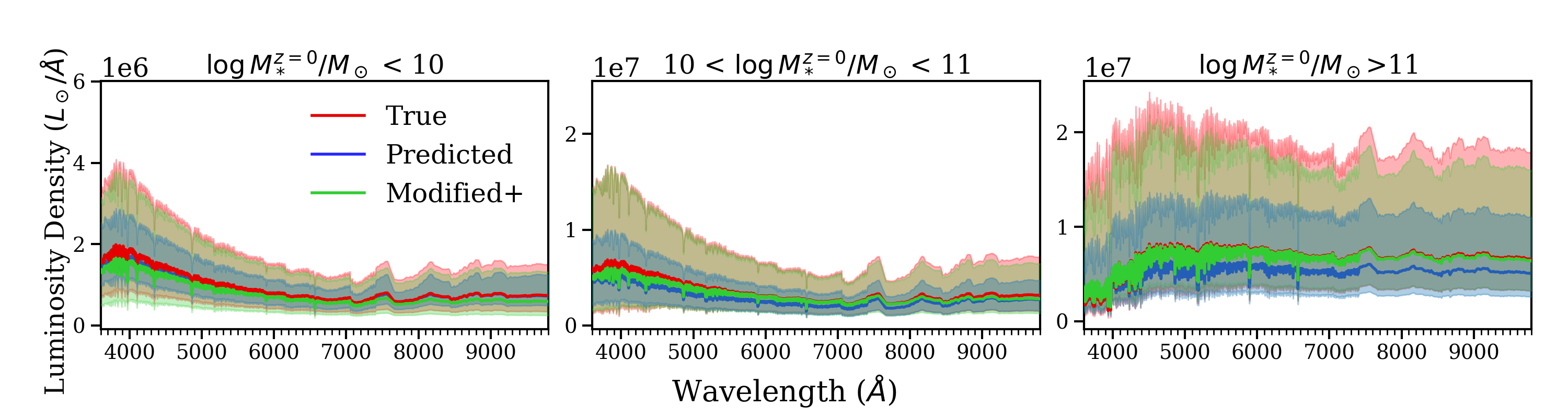}
    \caption{Spectral energy distributions (SEDs) for central galaxies where filter (iv) is excluded, results in a noticeable deterioration in the spectra for low-mass, star-forming galaxies, which show reduced accuracy compared to the results in Fig.\ref{fig:norm spectra} with the full set of filters.
}
    \label{fig: caseiv spectra}
\end{figure*}

\begin{figure}
    \includegraphics[width = \columnwidth]{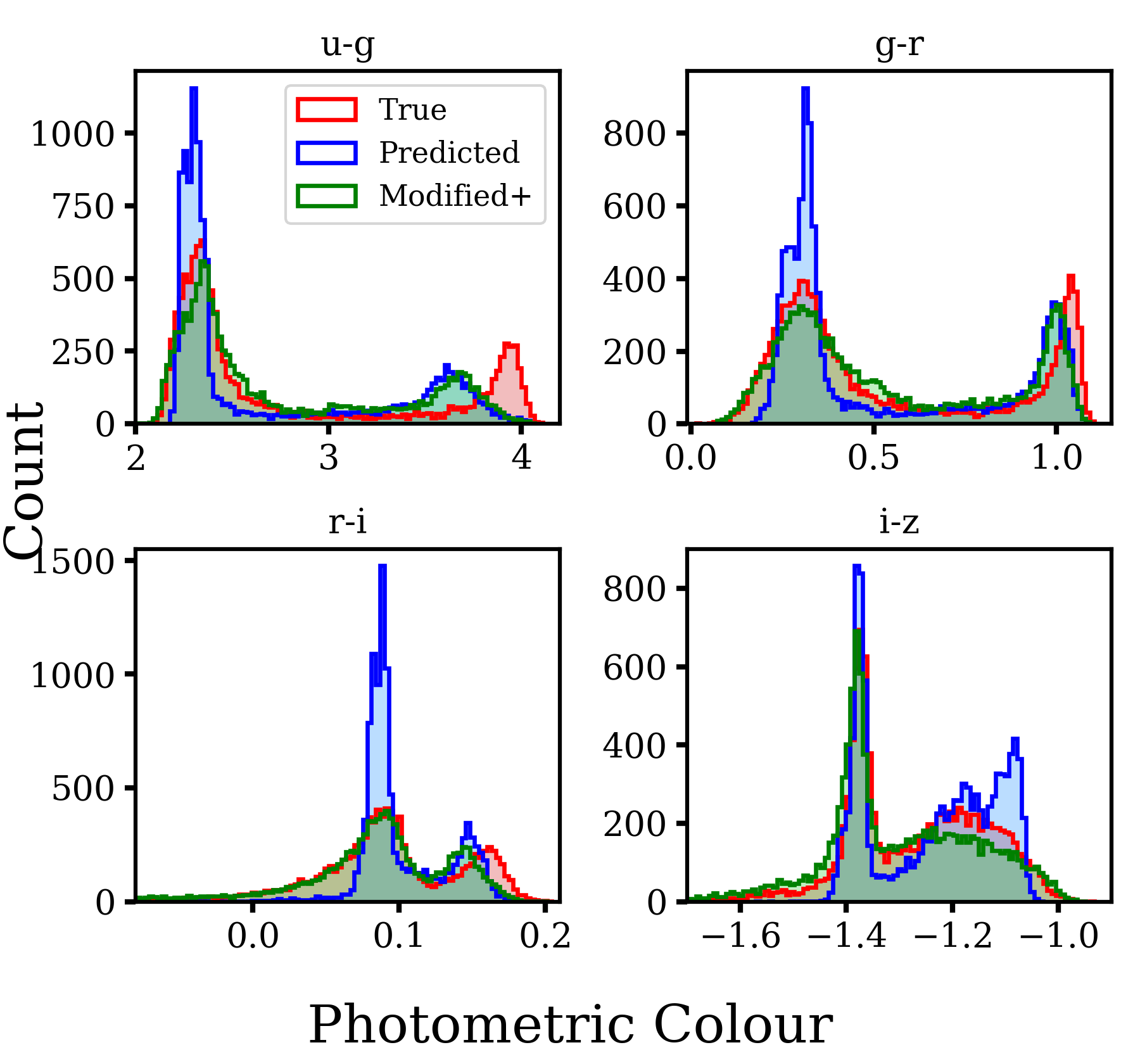}
    \caption{Color distributions for central galaxies where filter (iv) is excluded.  While the overall color recovery is effective, it improves further when this filter is included, as seen in Fig.\ref{fig:norm color} because the exclusion of this filter leads to slight discrepancies, particularly in blue, star-forming galaxies, where the scatter is not fully captured.
}
    \label{fig: caseiv color}
\end{figure}

Fig. \ref{fig: caseiv spectra} presents the spectra for central galaxies where filter (iv) is excluded. Compared to Fig. \ref{fig:norm spectra}, which includes this filter, the spectra for low-mass, star-forming galaxies are less accurate, showing a small but noticeable deterioration. This shortfall is also reflected in the color distribution (Fig. \ref{fig: caseiv color} compared to Fig. \ref{fig:norm color}), where excluding filter(iv) leads to a slight lack of correction for blue galaxies.

\section{Corrections in Metallicities}\label{sec:appendix-norm zh}
The stochastic corrections applied to individual metallicity histories (ZHs) are presented in Figure B1, which shows the ZHs for 15 random central galaxies. These corrections, similar to those applied to star formation histories (SFHs) in Fig. \ref{fig:cent norm sfh}, introduce short-timescale variability effectively, demonstrating the robustness of the method. By adding stochasticity, we are able to recover small-scale fluctuations in the ZHs that were not captured by the rNN predictions.

\begin{figure*}
    \includegraphics[width = \textwidth]{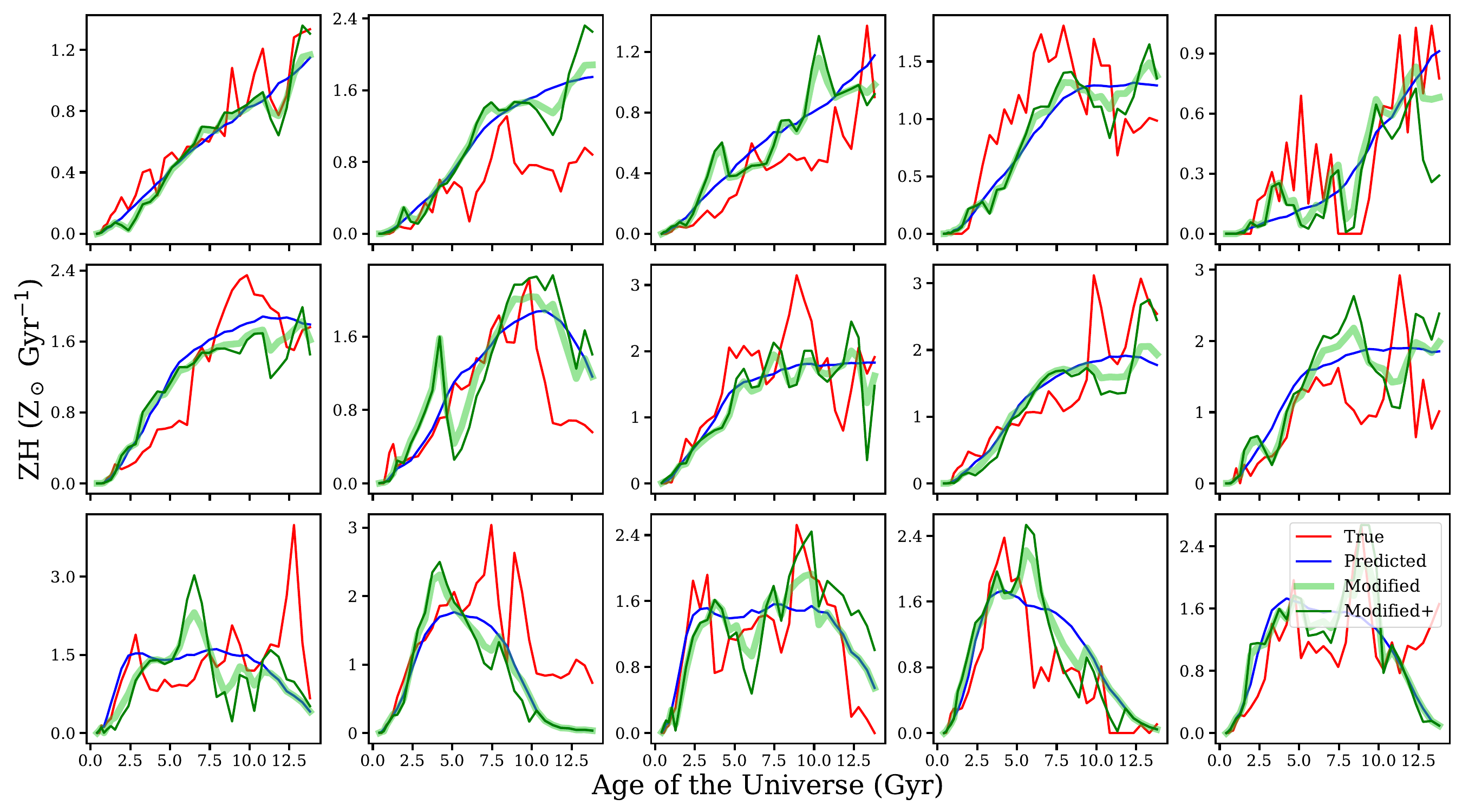}
    \caption{
Similar to Fig. \ref{fig:cent norm sfh}, but for the metallicity histories (ZHs) of 15 random central galaxies. The top panel shows low-mass galaxies ($\log M_*/M_\odot < 10$), the middle panel shows intermediate-mass galaxies ($10 \leq \log M_*/M_\odot < 11$), and the bottom panel shows high-mass galaxies ($\log M_*/M_\odot \geq 11$). Blue lines represent the smooth rNN-predicted ZHs, red lines show the true ZHs, light green lines show stochastically modified ZHs, and dark green lines show the results after normalization. Similar to SFHs \ref{fig:cent norm sfh} in Corrections effectively restore short-timescale variability in the ZHs as well.
}
    \label{fig:cent norm zh}
\end{figure*}
%%%%%%%%%%%%%%%%%%%%%%%%%%%%%%%%%%%%%%%%%%%%%%%%%%

% Don't change these lines
\bsp	% typesetting comment
\label{lastpage}
\end{document}